\documentclass[a4paper,twocolumn,11pt]{quantumarticle}
\pdfoutput=1
\usepackage[numbers,sort&compress]{natbib}
\usepackage[utf8]{inputenc}
\usepackage[english]{babel}
\usepackage[T1]{fontenc}
\usepackage{amsmath}
\usepackage{hyperref}
\usepackage{braket}
\usepackage{subcaption}
\usepackage{tikz}
\usepackage{lipsum}
\usepackage{physics}

\begin{document}

\title{Cat-Code-Protected Controlled Quantum Communication
via Non-Local CNOT Gates over Star Quantum Networks}

\author{Subham Das}
\affiliation{Instrumentation and Applied Physics,
Indian Institute of Science,
Bengaluru, India }
\email{subham1@iisc.ac.in}
\author{Abhishek Sharma}
\email{sharma.abhishek@cdac.in}
\affiliation{Centre for Development of Advanced Computing (C-DAC),
Chennai, India}
\author{Kailash S}
\email{kailashs@cdac.in}
\affiliation{Centre for Development of Advanced Computing (C-DAC),
Chennai, India}

\maketitle

\begin{abstract}
Controlled quantum communication enables secure state transfer between a sender and receiver with the assistance of one or more controllers. However, practical implementation over optical fibre networks is severely hindered by amplitude damping, which reduces fidelity exponentially with distance. We address this challenge by combining two powerful techniques: cat-state encoding for error correction and optimal non-local CNOT gates for distributed gate implementation. The protocol eliminates the need for the physical position qubit itself to travel through the optical fibre, reducing damping events. We show, through density-matrix simulations, that our cat-code-protected protocol with a non-local CNOT operation achieves higher fidelity at 50 km, significantly outperforming the standard protocol. We analyse the protocol's security against beam-splitter attacks and show that CHSH tests provide security against beam-splitter attacks on the distributed entanglement resource despite the cat code's error correction. Our results establish that cat-code-protected controlled quantum communication is feasible with current technology and structurally extensible to multiple controllers, providing a theoretical framework for studying error-corrected controlled quantum communication over long-distance star quantum networks.
\end{abstract}

\section{\label{SecI}Introduction}

Quantum networks constitute the fundamental infrastructure of the envisioned quantum internet, enabling the distribution of quantum information among spatially separated nodes through quantum channels. Such networks are expected to support a wide range of applications, including distributed quantum computing \cite{zhang2023federatedlearningquantumsecure}, secure quantum communication \cite{Pan_2024}, and quantum key distribution \cite{SINGH2025100089}. The realisation of these applications requires the reliable generation, distribution, and preservation of quantum coherence and entanglement across distant network nodes \cite{Kimble2008}. Consequently, the efficient implementation of quantum communication protocols over realistic network architectures has become a fundamental challenge in the development of the quantum internet.

Controlled quantum communication is an important primitive in quantum networks because it enables secure transmission of quantum information only with the cooperation of one or more controllers. For example, in the work \cite{Ting_2005}, a controlled quantum communication with one controller is done with a triplet state. Recent studies have also investigated its implementation over realistic quantum networks. For instance, Yan et al. \cite{Yan:24} proposed a remote-controlled quantum teleportation network based on polarisation squeezed states, demonstrating the feasibility of controlled teleportation among multiple remote users over optical communication channels. Despite these advances, mitigating photon-loss-induced amplitude damping and implementing controlled quantum communication over realistic star quantum networks remain open challenges.

 When transferring a quantum state through a quantum network, the state degrades due to environmental interactions, introducing noise. One of the dominant noise mechanisms in optical-fibre communication is amplitude damping. Several approaches have been proposed to mitigate amplitude damping in quantum communication. Decoherence Free subspaces (DFS) \cite{PhysRevLett.81.2594} protect quantum information by encoding it into a noise-invariant subspace. An existing work, such as \cite{Qin2009}, in which the computational state is encoded into DFS states under collective amplitude damping. A similar work \cite{GuedesAssis2013}, which discusses the implementation of the BB84 protocol \cite{BENNETT20147} in a medium with collective amplitude damping. Although these methods effectively suppress collective noise, they depend on specific noise symmetries. They thus are not directly applicable to general amplitude damping, which occurs in long-distance optical fibre networks.

An alternative approach is to use quantum error-correcting codes specifically designed for amplitude damping. Four-qubit code \cite{PhysRevA.56.2567}, and the smallest quantum code for amplitude damping \cite{smallestqcam} is such an example, which can correct single-amplitude-damping errors. However, if multiple photon losses occur, this method is not scalable. Along with that, there are cat codes \cite{PhysRevA.70.022317}. These methods convert or correct photon-loss errors, thereby significantly improving communication fidelity. However, most existing studies focus on point-to-point quantum communication or quantum memories. Their integration with distributed controlled quantum communication protocols, particularly those implemented over realistic star quantum networks using non-local quantum gates \cite{PhysRevA.62.052317}, has received comparatively little attention.

In this work, we investigate the implementation of controlled quantum communication over a realistic star quantum network. We replace the repeated transmission of the position qubit with distributed non-local CNOT operations, thereby reducing the effect of photon loss during communication. Furthermore, cat-state encoding is employed during entanglement distribution to convert amplitude-damping errors into correctable logical phase-flip errors, improving the communication fidelity over long distances. Finally, we analyse the security of the proposed protocol against beam-splitter attacks under ideal channel conditions with cat-state encoding. The present work focuses on the effect of optical-fibre amplitude damping. Other imperfections, such as gate errors, memory decoherence, detector inefficiency, and imperfect syndrome extraction, are not considered. Also, the effect of fibre loss on the cat qubit is modelled at the logical level through the induced phase-flip probability while the cat-state amplitude is kept fixed throughout the protocol.

The main contributions of this work are as follows:
\begin{itemize}
    \item Implementation of controlled quantum communication over a star quantum network.
\item Replacement of repeated position-qubit transmission using optimal non-local CNOT gates.
\item Integration of cat-state encoding for mitigating amplitude-damping errors.
\item Comparative numerical evaluation of four communication architectures.
\item Security analysis of the proposed protocol against beam-splitter attacks.
\end{itemize}
The remainder of the paper is organised as follows. Section \ref{prel} presents the preliminaries. Section \ref{sec:level3} describes the theoretical framework of the proposed algorithms. Section \ref{sec4} presents the implementation of the controlled quantum communication protocol in a network. Section \ref{sec6} provides the security analysis of the protocol. Section \ref{sec7} presents the simulation results. Section \ref{sec8} discusses the findings and concludes the paper. Section \ref{sec9} discusses the future works.

\section{\label{prel}Preliminaries}


\subsection{Noise model}
Noise refers to a disturbance that affects the operation of quantum devices and the accuracy of quantum information processing. Noise can be modelled using a completely trace-preserving (CPTP) map called the Kraus operators. It satisfies the completeness condition given by $\sum_{i=0}^{N} E_{i}^{\dagger}  E_{i}=I$. Here, $E_{i}$ are the Kraus operators. The completeness condition for Kraus operators, $\sum_i E_i^\dagger E_i = I$, ensures that the evolved state $\rho'$ is normalised and has a unit trace, which are the properties required for a valid quantum state. A Kraus operator can be used to describe how a quantum operation or channel affects a quantum state. Given a quantum state $\rho$ and a set of Kraus operators ${E_i}$ that describe the effect of the quantum operation or channel, the evolved state after the operation is given by:

\begin{equation}
    \rho' = \sum_{i=0}^{N} E_{i} \rho E_{i}^{\dagger}
    \label{kraus}
\end{equation}

The Kraus operators act on the quantum state $\rho$ in a linear fashion, and the resulting state $\rho'$ is a weighted sum of the individual terms of the form $E_i \rho E_i^\dagger$. Each term in this sum represents the effect of a particular Kraus operator on the initial state, and the weights of these terms depend on the probability amplitudes of the quantum state.
\subsubsection{Phase flip}
A phase flip error occurs when the phase of a qubit is unintentionally flipped. The Kraus operator for phase flip is given by,
\begin{equation}
\begin{split}
E_0 &= \sqrt{1 - p} \, I, \\
E_1 &= \sqrt{p} \, Z.
\end{split}
\end{equation}

\subsubsection{Amplitude Damping}
It arises from energy loss in the system. For amplitude damping, the Kraus operators are given by \begin{equation}
\begin{split}
E_0 &= \ket{0}\bra{0} + \sqrt{1 - p} \ket{1}\bra{1}\\
E_1 &= \sqrt{p} \ket{0}\bra{1} 
\end{split}
\end{equation}
For an optical fibre,
\begin{equation}
    p=1-\eta
\end{equation}
Where $\eta$ denotes the transmission of the optical fibre. It is given by
\begin{equation}
    \eta = \exp(-\kappa L),
\end{equation}
where $\kappa = 0.04605~\mathrm{km}^{-1}$ is the attenuation coefficient of the optical fibre and $L$ is the transmission distance.

\section{Theoretical framework for controlled quantum communication \label{sec:level3} }
\subsection{Overview of the communication protocol\label{pre1}}

We consider a quantum communication protocol based on a discrete-time quantum walk framework, as presented in \cite{Das2025}, involving four subsystems: the sender (coin space), the position space, the controllers (coin space), and the receiver (coin space). The goal of the protocol is to transfer an arbitrary quantum state from the sender to the receiver through controlled unitary operations and measurement.
\begin{itemize}
    
\item The sender initially prepares an unknown quantum state
\begin{equation}
|\psi\rangle = \alpha |0\rangle + \beta |1\rangle,
\end{equation}
which resides in the sender's coin space. The position space and receiver coin space are initialised to $\ket{0}$.

\item First, the sender applies a CNOT operation, using the coin space as control and the position space as the target. This step entangles the sender's state with the position degree of freedom, effectively distributing quantum information across the system.

\item In the next step, each controller performs a Hadamard operation on its own qubit and applies a CNOT operation on the position qubit. This step entangles the controller's qubit with the position qubit.

\item Similarly to the controller, the receiver applies a Hadamard operation on its coin space, followed by a CNOT operation on the position with its coin space as the controller. 

\item Finally, the sender applies a Hadamard operation on its coin space and measures the state in the computational basis. The measurement outcome is classically communicated to the receiver, who then applies a corresponding correction operation to recover the original state.
\end{itemize}
In the absence of noise, the protocol allows faithful reconstruction of the quantum state at the receiver. However, in realistic scenarios, noise processes such as bit-flip, phase-flip, and amplitude damping errors degrade the fidelity of transmission. Thus, in this work, we are discussing the effect of cat-state encoding with repetition encoding, which can protect the state from amplitude damping. We will refer to this combination as an error-correcting cat qubit.

\subsection{Cat-state encoding}

 A cat state encoding encodes the states to coherent states :
\begin{equation}
|C_{\pm}\rangle = \ket{\pm \alpha}
\end{equation}

These states form a logical qubit basis:
\begin{equation}
|0_L\rangle = |C_{+}\rangle, \quad |1_L\rangle = |C_{-}\rangle.
\end{equation}
The unique property of this cat state encoding is that it converts the amplitude damping error into phase flip error  \cite{PhysRevA.81.062344}.
\subsubsection{Photon loss error}
A prominent advantage of cat-state encoding is that it converts the photon-loss problem into a phase-flip error, which is a unitary error; thus, stabiliser codes can be used to correct it. The photon loss action on the cat states is as follows:
\begin{equation}
a |C_{\pm}\rangle \propto \pm|C_{\pm}\rangle,
\end{equation}
 Thus, this encoding effectively converts amplitude damping, a non-Pauli error, into an effective Pauli-$Z$ error.

If the coherent amplitude, the overlap between $|\alpha\rangle$ and $|-\alpha\rangle$, becomes negligible, the mapping to a phase-flip error becomes increasingly accurate. 

This transformation of continuous bosonic noise into a predominantly discrete error model enables the application of standard qubit-based error correction codes, such as repetition codes.

\subsubsection{Error Analysis}

Cat-state encoding alone cannot correct photon-loss errors. It transforms the amplitude-damping channel into an effective phase-flip error in the logical basis. To suppress these logical phase-flip errors, a second layer of encoding using the repetition code is implemented. The logical codewords are

\begin{align}
|0_L\rangle &= \ket{C_{+}}^{\otimes n},\\
|1_L\rangle &= \ket{C_{-}}^{\otimes n},
\end{align}

The $n$-qubit repetition code can correct up to $t$ phase-flip errors, where

\begin{equation}
t=\left\lfloor\frac{n-1}{2}\right\rfloor,
\end{equation}

thereby significantly reducing the effective logical phase-flip error probability. The probability that the repetition code fails, corresponding to the occurrence of more than $t$ phase-flip errors, is given by

\begin{equation}
P_{\mathrm{fail}}
=
\sum_{k=t+1}^{n}
\binom{n}{k}
p_e^{k}
\left(1-p_e\right)^{\,n-k},
\label{t_qubit}
\end{equation}
where $p_e$ is given as,
\begin{equation}
p_e
=
\frac{1-e^{-2(1-\eta)\alpha^2}}{2},
\label{eq:pz}
\end{equation}

where $\eta$ is the channel transmissivity and $\alpha$ is the coherent-state amplitude. This expression is derived from the work \cite{PhysRevA.81.062344} Thus, there is a trade-off between the distinguishability of the state and the effect of noise on the state.

Equation~\eqref{eq:pz} can also be inverted to determine the coherent-state amplitude required to achieve a desired logical phase-flip probability,

\begin{equation}
\alpha
=
\sqrt{
\frac{-\ln\!\left(1-2p_e\right)}
{2(1-\eta)}
}.
\label{eq:alpha_pz}
\end{equation}

Equation~\eqref{eq:alpha_pz} provides a direct guideline for selecting the cat-state amplitude to maintain a target logical error rate for a given channel transmissivity. Moreover, the cat's logical state has a finite overlap given by
\begin{equation}
    O_1 = \langle -\alpha | \alpha \rangle
    = \exp\left(-2|\alpha|^2\right).
\end{equation}

Thus, however, giving a non-zero overlap between states on increasing the logical code to N repetitions. The overlap between the logical states decreases, given by
\begin{equation}
    O_N=O_1^N
\end{equation}
If N is sufficiently large, the overlap between two logical states goes to zero.

\section{Implementation of Controlled Quantum Communication in Star Networks\label{sec4}}
The implementation of the protocol can take multiple forms in a star network; one such implementation is to consider direct state transfer of the position qubit. Another such method is to use non-local optimal CNOT operations with a Bell state, which involves applying a CNOT operation between two qubits via one EPR pair and 2 classical bits. One can improve the fidelity of such a protocol by using error-correction methods, such as error-correcting cat qubit encoding. 
\subsection{Direct Quantum Walk-Based Implementation}

Implementing controlled quantum communication in a star network requires the position qubit to be transmitted through optical fibre links connecting the participating nodes. In the standard controlled teleportation protocol, the position qubit undergoes multiple transmissions during the communication process. Specifically, it travels.

\begin{itemize}
\item from the service provider to the sender
\item from the sender to the service provider,
\item from the service provider to each controller,
\item from each controller back to the service provider,
\item from the service provider to the receiver, 
\item from the receiver back to the service provider, and
\item from the service provider to the sender.
\end{itemize}

\begin{figure}
    \centering
    \includegraphics[width=0.8\linewidth]{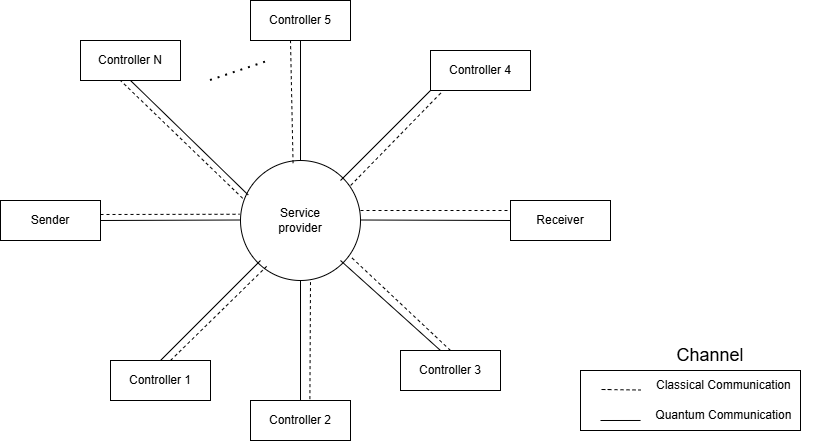}
    \caption{Schematic diagram of the controlled quantum communication protocol implemented in a star network. The Service Provider performs error correction on the travelling qubits. Classical communication channels are used for exchanging measurement results. }
    \label{fig:schematicfig}
\end{figure}

Each transmission through an optical fibre is subject to photon loss, which is well described by the amplitude damping channel. Consequently, the position qubit undergoes multiple independent amplitude-damping events throughout the protocol. For $M$ controllers, the position qubit undergoes $2M+5$ amplitude damping in an optical fibre of length L. Since transmission fidelity decreases exponentially with fibre length, repeated transmissions lead to cumulative degradation of the quantum state, thereby significantly reducing the overall fidelity of the controlled quantum communication protocol. The figure \ref{fig:schematicfig} shows a schematic diagram of the architecture of the quantum communication protocol. The protocol performance can be further improved by replacing a standard position qubit with an error-correcting cat qubit encoded in position, and, for n-repetition codes, it corrects t errors before transmission through each optical fibre. Even the service provider is involved in the error correction process.

\subsection{Limitations of the Direct Architecture}

In the direct architecture model, since the position qubit must physically traverse the optical fibre network multiple times, it experiences repeated amplitude damping during transmission. For a protocol involving $M$ controllers, the position qubit undergoes several independent fibre transmissions, including communication between the sender, service provider, controllers, and receiver. Consequently, the accumulated photon loss increases with both transmission distance and the number of controllers, leading to an exponential degradation in the fidelity of the reconstructed state.

Although the direct implementation realises the quantum walk protocol, its dependence on repeated transmission of the position qubit makes it unsuitable for long-distance quantum communication. This limitation motivates the replacement of the travelling position qubit by distributed non-local quantum gates, as discussed in the following subsection.

\subsection{Cat-Code-Protected Non-Local CNOT Architecture}
The original protocol used direct transmission of the position qubit, allowing each member to apply a CNOT operation. This introduces noise to the protocol. By replacing each physical transmission of a position qubit with a non-local CNOT, we:
\begin{enumerate}
\item Eliminate the physical transmission of the position qubit for the CNOT operation with the non-local CNOT operation.
\item For each non-local CNOT, only the Bell-pair resource is distributed through the optical-fibre network. The travelling qubit of the Bell pair is therefore exposed to amplitude damping during its transmission from the sender to the service provider and subsequently from the service provider to the intended party.
\item  The number of amplitude-damping events is reduced from $2M+5$ in the direct-transmission architecture to $2M+2$ in the non-local-CNOT architecture. 
\item Moreover, the damping no longer acts only on the position qubit through repeated transmission. Instead, the optical loss affects the distributed Bell pairs used to implement the non-local operations. This becomes increasingly important as the communication distance increases.
\end{enumerate}

The sender, each controller, and the receiver each perform a non-local CNOT with the receiver's ancilla qubits. In this way, the position degree of freedom is implemented virtually through non-local operations rather than by physically routing the position qubit between nodes. To further suppress the error, the Bell states are encoded using an error-corrected cat qubit, thereby improving performance.

This non-local CNOT with an error-correcting cat qubit Bell state thus serves as the fundamental building block for our cat-code-protected controlled communication protocol, enabling higher-fidelity state transfer over long distances.

In the numerical analysis, we consider the heralded successful branch of the non-local CNOT, in which the required measurement outcomes are zero. Trials yielding any other measurement outcome are discarded. The reported fidelity, therefore, corresponds to the conditional fidelity of the successfully heralded operation.

\section{Security Analysis\label{sec6}}
In this section, we discuss the security of the cat-code-protected state when implemented via a non-local CNOT protocol, and analyse it against eavesdropping attacks. We consider an adversary Eve who can tap the optical fibre and extract information about the transmitted state using a beam splitter. We show that while cat codes protect against errors, they introduce unique security vulnerabilities that require careful analysis. To analyse the security of the protocol independently of intrinsic channel loss.

\subsection{Adversary Model}

We consider an eavesdropper Eve with the following capabilities:
\begin{enumerate}
\item Eve can tap the optical fibre using a beam splitter, diverting a fraction of the light to her own measurement apparatus.
\item Eve can measure the tapped light to extract information about the transmitted cat state.
\item Eve cannot access the local operations at the sender, controllers, or receiver nodes.
\item Eve's goal is to learn information about the transmitted quantum state without being detected.
\end{enumerate}

This model is realistic for optical fibre networks, where beam splitters are passive optical components that can be inserted into the fibre with minimal disturbance.

Cat states are coherent superpositions of two coherent states with opposite amplitudes:

\begin{equation}
|\psi\rangle = a|\alpha\rangle + b|-\alpha\rangle
\label{eq:cat_state}
\end{equation}

A key property of coherent states is that they are eigenstates of the annihilation operator:

\begin{equation}
\hat{a}|\alpha\rangle = \alpha|\alpha\rangle, \qquad \hat{a}|-\alpha\rangle = -\alpha|-\alpha\rangle
\label{eq:eigenstates}
\end{equation}

When Eve taps the fibre with a beam splitter of transmissivity $T$ (where $1-T$ is the fraction tapped by Eve), the coherent state transforms as:

\begin{equation}
|\alpha\rangle \ket{0} \xrightarrow{\text{BS}} |\alpha\sqrt{T}\rangle \otimes |\alpha\sqrt{1-T}\rangle
\label{eq:bs_alpha}
\end{equation}

\begin{equation}
|-\alpha\rangle\ket{0} \xrightarrow{\text{BS}} |-\alpha\sqrt{T}\rangle \otimes |-\alpha\sqrt{1-T}\rangle
\label{eq:bs_minus_alpha}
\end{equation}

\begin{figure}
    \centering
    \includegraphics[width=0.8\linewidth]{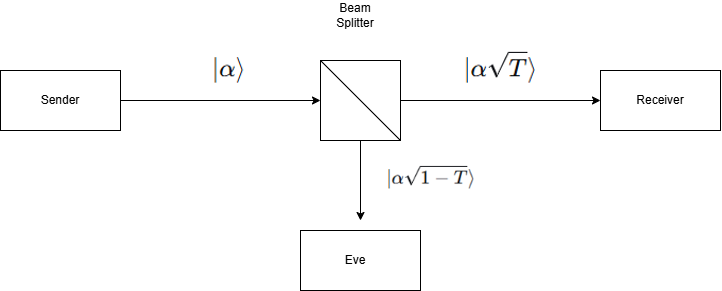}
    \caption{Schematic of Eve's beam-splitter attack on the travelling cat state. The coherent state $\ket{\alpha}$ enters a beam splitter of transmissivity $T$. The main beam $\ket{\alpha\sqrt{T}}$ continues to the intended receiver (Bob), while the tapped beam $\ket{\alpha\sqrt{1-T}}$ is diverted to Eve's measurement apparatus. Because coherent states are eigenstates of the annihilation operator, the superposition is preserved on both beams, allowing Eve to extract information without collapsing the quantum state.}
    \label{Beamsplitterattack}
\end{figure}

The action of the beam splitter is shown in Figure \ref{Beamsplitterattack}. Therefore, the full cat state after Eve's beam splitter becomes:

\begin{equation}
\begin{aligned}
(a|\alpha\rangle + b|-\alpha\rangle)\ket{0}
&\xrightarrow{\mathrm{BS}}
a|\alpha\sqrt{T}\rangle|\alpha\sqrt{1-T}\rangle \\
&\qquad + b|-\alpha\sqrt{T}\rangle|-\alpha\sqrt{1-T}\rangle
\end{aligned}
\label{eq:bs_cat}
\end{equation}

This action is equivalent to a CNOT operation on a qubit. Thus, the attack on the Bell state would produce a GHZ state :

\begin{equation}
\begin{aligned}
&\frac{1}{\sqrt{2}}
\left(\ket{0}\ket{\alpha}+\ket{1}\ket{-\alpha}\right)\ket{0}
\xrightarrow{\mathrm{BS}}\\
&\frac{1}{\sqrt{2}}\Big(
\ket{0}\ket{\alpha\sqrt{T}}\ket{\alpha\sqrt{1-T}}\\
&\qquad
+\ket{1}\ket{-\alpha\sqrt{T}}\ket{-\alpha\sqrt{1-T}}
\Big).
\label{eq:ghz_state}
\end{aligned}
\end{equation}

 This state can be used to design an attack to steal information at the latter state, as shown in the next section.

\subsection{Attack on the Entanglement Resource of the Non-Local CNOT}

The implementation of a non-local CNOT requires Alice and Bob to share an entangled Bell pair that acts as the resource state for the distributed gate. While the data qubits remain stored locally at their respective nodes, one half of the Bell pair must be distributed through the quantum network. Consequently, an adversary may attempt to interact with the distributed resource before the non-local CNOT protocol consumes it.

Consider Alice and Bob sharing the Bell state
\begin{equation}
|\Phi^+\rangle_{A_1B_1}
=
\frac{1}{\sqrt{2}}
\left(
|00\rangle_{A_1B_1}
+
|11\rangle_{A_1B_1}
\right),
\end{equation}
where $A_1$ and $B_1$ denote Alice's and Bob's resource qubits, respectively. During transmission through the optical fibre of the qubit $B_1$ from Alice to Bob, Eve can perform a beam splitter attack on the qubit. It interacts with the transmitted resource qubit $B_1$ through a beam splitter. This action is equivalent to a CNOT operation. The joint resource state then transforms according to
\begin{equation}
|\Phi^+\rangle_{A_1B_1}|0\rangle_E
\longrightarrow
\frac{1}{\sqrt{2}}
\left(
|000\rangle_{A_1B_1E}
+
|111\rangle_{A_1B_1E}
\right).
\end{equation}

Thus, Eve converts the bipartite Bell resource into a tripartite GHZ state,
\begin{equation}
|\mathrm{GHZ}\rangle_{A_1B_1E}
=
\frac{1}{\sqrt{2}}
\left(
|000\rangle+|111\rangle
\right).
\end{equation}

\begin{figure}
    \centering
    \includegraphics[width=1.0\linewidth]{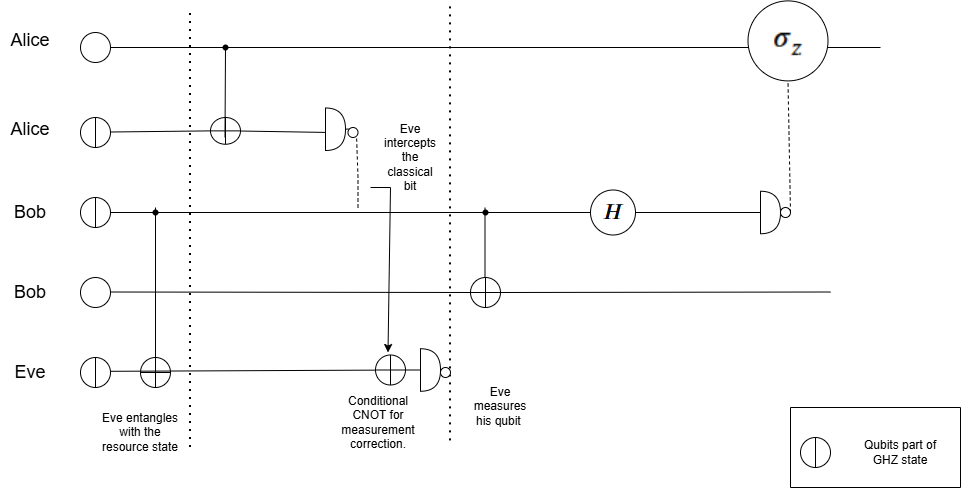}
    \caption{Schematic of Eve's attack on the entanglement resource of the non-local CNOT. Eve couples an ancillary qubit to the distributed Bell pair through a local CNOT, creating a tripartite GHZ state. The subsequent local operation and measurement by Alice correlate Eve's ancilla with Alice's information qubit.}
    \label{fig:placeholder}
\end{figure}

This state forms the initial resource state, which Eve will use to attack to access the information during the non-local CNOT  operation. The figure \ref{fig:placeholder} gives a diagrammatic explanation of the attack design:
\begin{equation}
|\psi\rangle_A
=
a|0\rangle_A+b|1\rangle_A,
\qquad
|a|^2+|b|^2=1.
\end{equation}

The combined state prior to the non-local CNOT is therefore
\begin{equation}
|\Psi\rangle
=
\left(
a|0\rangle_A+b|1\rangle_A
\right)
\otimes
\frac{1}{\sqrt{2}}
\left(
|000\rangle+|111\rangle
\right)_{A_1B_1E}.
\end{equation}

According to the non-local CNOT protocol, Alice first performs a local CNOT operation between her information qubit $A$ and her resource qubit $A_1$,
\begin{equation}
\operatorname{CNOT}_{A\rightarrow A_1}.
\end{equation}

After this operation, the joint state becomes
\begin{align}
|\Psi'\rangle
=
\frac{1}{\sqrt{2}}
\Big[
&a|0\rangle_A
\left(
|000\rangle+|111\rangle
\right)_{A_1B_1E}
\nonumber\\
+&
b|1\rangle_A
\left(
|100\rangle+|011\rangle
\right)_{A_1B_1E}
\Big].
\end{align}

Alice then measures $A_1$ in the computational basis, as required by the non-local CNOT protocol.

If Alice obtains the measurement outcome $m_A=0$, the remaining state becomes
\begin{equation}
|\Psi_{m_A=0}\rangle
=
a|0\rangle_A|0\rangle_{B_1}|0\rangle_E
+
b|1\rangle_A|1\rangle_{B_1}|1\rangle_E.
\end{equation}

On the other hand, if Alice obtains $m_A=1$, the remaining state is
\begin{equation}
|\Psi_{m_A=1}\rangle
=
a|0\rangle_A|1\rangle_{B_1}|1\rangle_E
+
b|1\rangle_A|0\rangle_{B_1}|0\rangle_E.
\end{equation}

The measurement outcome $m_A$ must be communicated classically to Bob so that he can perform the appropriate feed-forward correction required by the non-local CNOT. If this classical communication is publicly accessible, Eve can also learn $m_A$. She therefore knows whether her ancilla is directly or inversely correlated with Alice's computational-basis state.

If the classical bit shared from Alice to Bob is $m_A=0$, then Eve does not need to do any further operation and can directly measure his qubit in the computational basis, which gives
\begin{equation}
P(E=0)=|a|^2,
\qquad
P(E=1)=|b|^2.
\end{equation}

If the classical bit Alice shares with Bob is $m_A=1$, the association is reversed. However, because the classical bit is publicly communicated. Eve can correctly interpret it. Equivalently, she may conditionally apply
\begin{equation}
X_E^{m_A}
\end{equation}
to place her ancilla in a common correlation convention before measurement.

Therefore, through this series of attacks, by first becoming entangled with the Bell resource and subsequently exploiting the classical bit information shared from Alice to Bob of the non-local CNOT protocol, Eve can obtain information about the computational-basis populations of Alice's information qubit. This attack does not violate the no-cloning theorem, since Eve never creates an independent copy of the arbitrary state $a|0\rangle+b|1\rangle$. Rather, information leakage arises from correlations established through a compromised entanglement resource.

It is important to understand that a single measurement does not allow Eve to reconstruct an arbitrary unknown quantum state. In particular, the above attack does not reveal the relative phase between $a$ and $b$. Nevertheless, it demonstrates that the integrity of the entangled resource used for the non-local CNOT is itself a security requirement. This motivates verification of the distributed Bell pairs before they are consumed in the communication protocol.

\subsection{CHSH Test for the Cat-Encoded GHZ State}

To detect Eve's presence, Alice and Bob can perform CHSH tests. The state in Eq.~\ref{eq:ghz_state} is a GHZ-like entangled state shared among Alice, Bob, and Eve. When Alice and Bob trace out Eve's system, the reduced state has coherence:

\begin{equation}
\mathcal{O}_C = \langle -\alpha\sqrt{1-T} | \alpha\sqrt{1-T} \rangle = e^{-2(1-T)|\alpha|^2}
\label{eq:oc}
\end{equation}

The maximum CHSH value for this state is:

\begin{equation}
\boxed{
S_{\text{max}} = 2\sqrt{1 + \mathcal{O}_C^2}
}
\label{eq:chsh_result}
\end{equation}

Substituting Eq.~\ref{eq:oc}:

\begin{equation}
S_{\text{max}} = 2\sqrt{1 + e^{-4(1-T)|\alpha|^2}}
\label{eq:chsh_alpha}
\end{equation}

The security of the protocol depends on the parameter:

\begin{equation}
x = (1-T)|\alpha|^2
\label{eq:x_param}
\end{equation}

\subsection{Eve's Distinguishability}

Eve's optimal probability of correctly distinguishing between the two coherent states
is given by the Helstrom bound:

\begin{equation}
P_{\mathrm{correct}}
=
\frac{1}{2}
\left(
1+\sqrt{1-e^{-4x}}
\right),
\label{eq:helstrom}
\end{equation}

where

\[
x=(1-T)|\alpha|^2.
\]

Let us define Eve's \emph{distinguishability} $\epsilon$ as the excess over random guessing:

\begin{equation}
\epsilon
=
P_{\mathrm{correct}}
-
\frac{1}{2}.
\end{equation}

Substituting Eq.~\eqref{eq:helstrom},

\begin{equation}
\epsilon
=
\frac{1}{2}
\sqrt{1-e^{-4x}}.
\label{eq:epsilon_def}
\end{equation}

Solving for $x$ in terms of $\epsilon$,

\begin{align}
2\epsilon
&=
\sqrt{1-e^{-4x}},\\
4\epsilon^2
&=
1-e^{-4x},\\
e^{-4x}
&=
1-4\epsilon^2.
\end{align}

Taking the natural logarithm,

\begin{equation}
x
=
-\frac{1}{4}
\ln\!\left(1-4\epsilon^2\right)
\label{eq:x_from_epsilon}
\end{equation}

This gives the amount of light Eve must tap (quantified by $x$) to achieve a distinguishability $\epsilon$.

\subsection{Statistical Detection Threshold}

For $N$ CHSH testing rounds, Alice and Bob can detect Eve only if the drop in the CHSH value is larger than the statistical uncertainty. Since each round is independent, we can denote each round as $S_1, S_2, S_3... S_{N_{exp}}$, then by the central limit theorem, the uncertainty in the estimated CHSH value scales as:

\begin{equation}
\delta = \frac{\sigma_0}{\sqrt{N_{exp}}}=\frac{1}{\sqrt{N_r}}
\end{equation}

Thus, Eve is detected when:

\begin{equation}
S_{\text{max}} < 2\sqrt{2} - \frac{1}{\sqrt{N_r}}
\label{eq:detection_threshold}
\end{equation}

Thus, the critical value of $x$ above which Eve is detected is:

\begin{equation}
\boxed{x_{\text{crit}} = \frac{1}{2\sqrt{2N_r}}}
\end{equation}
If the tapping by Eve is greater than $x_{\text{crit}}$, an $N_{exp}$-round CHSH test will detect it. Using Eq.~\ref{eq:x_from_epsilon}, we can deduce that Eve is detected when:

\begin{equation}
-\frac{1}{4}\ln(1 - 4\epsilon^2) > \frac{1}{2\sqrt{2N_r}}
\end{equation}

Thus, when the distinguishability of the Eve state is beyond the threshold \ref{threshold}, the state will be 
\begin{equation}
\epsilon > \frac{\sqrt[4]{2}}{2\sqrt[4]{N_r}}=\epsilon_{crit}
\end{equation}

So the detectable distinguishability scales as
\begin{equation}
    \epsilon_{crit}\propto N_r^{-1/4}
\end{equation}

Thus, as the number of rounds increases, Eve's distinguishability decreases, thereby quantifying her information gain.

 Using $\epsilon_{crit}$ as an upper bound of the Eve's cat qubit state, the cat value can be estimated. The overlap between the coherent states is $|\beta\rangle$ and $|-\beta\rangle$. Consequently, the coherent-state amplitude can be estimated directly from Eve's distinguishability using equation \ref{eq:x_from_epsilon} given as

\begin{equation}
|\beta|^2
=
-\frac{1}{4}
\ln\!\left(1-4\epsilon_{crit}^2\right).
\label{eq:beta_from_epsilon}
\end{equation}

If Eve taps the communication channel using a beam splitter of transmissivity $T$, then the tapped coherent-state amplitude satisfies.

\begin{equation}
|\beta|^2=(1-T)|\alpha|^2,
\label{abc}
\end{equation}

where $|\alpha\rangle$ is the coherent state transmitted by Alice. Therefore, by increasing the number of trials $N_r$, we decrease the distinguishability $\epsilon$, which also weakens Eve's attack by increasing transmissivity. Hence, Eve's beam transmissivity can be estimated as

\begin{equation}
T\geq 1+
\frac{1}{4|\alpha|^2}
\ln\!\left(1-4\epsilon_{crit}^2\right)
=
1-
\frac{-\ln\!\left(1-4\epsilon_{crit}^2\right)}
{4|\alpha|^2}.
\label{eq:T_from_epsilon}
\end{equation}

Thus, even if Eve is present, its information gain can be controlled by choosing $N_r$ sufficiently large so that $\epsilon$ is arbitrarily small. Thus, increasing $N_r$ makes a weaker attack detectable.

\section{Implementation and Results\label{sec7}}

In this section, we present numerical results from density-matrix simulations of our cat-code-protected controlled quantum communication protocol. We compare four protocols: (i) the standard protocol with no encoding, (ii) the cat-code-protected protocol, (iii) the non-local CNOT protocol, and (iv) the combined cat-code-protected non-local CNOT protocol. We analyse fidelity as a function of distance, the effect of the cat code amplitude $\alpha$, and the effect of the repetition code. To isolate the effect of amplitude, the following assumptions are adopted:
\begin{itemize}
    \item quantum gates are ideal.
    \item State preparation and measurements are assumed to be perfect.
    \item Syndrome extraction and error correction are ideal.
    \item Quantum memories are assumed to have infinite coherence time.
    \item Classical communication is assumed to be noiseless.
    \item Amplitude damping is the only source of noise considered, which arises from the optical fibre.
    \item The coherent-state amplitude is kept fixed, and the attenuation-induced logical phase-flip probability is the only effect of amplitude damping included in the simulation.
\end{itemize}
\subsection{Numerical Implementation}

To compare the performance of the proposed communication schemes, we developed a density-matrix simulation framework that implements four protocols under identical channel conditions. The first implementation corresponds to the original controlled quantum communication protocol, in which the position qubit is physically transmitted through the optical fibre connecting the nodes of the star network. The second implementation replaces the position qubit with a single phase-flip-error-correcting cat qubit, in which amplitude damping is mapped to an equivalent logical phase-flip channel. The logical phase-flip probability is evaluated using Eqs.~(\ref{t_qubit}) and (\ref{eq:alpha_pz}), while the encoding, syndrome extraction, correction, and decoding operations are assumed to be ideal. Consequently, the simulation provides an upper bound on the performance achievable with the error-correcting cat code.

The third implementation replaces repeated position-qubit transmissions with optimal non-local CNOT gates implemented using shared Bell pairs. Finally, the fourth implementation combines the non-local CNOT architecture with error-correcting cat-state encoding by assuming that the distributed Bell pair is encoded using the same logical error model described above. The measurement outcomes are assumed to correspond to the branch in which no feed-forward correction is required.

For all simulations performed, the star network is assumed to be symmetric, with every party located at an equal distance \(L\) from the service provider. Photon loss in each optical link is modelled independently using the amplitude-damping channel. Before each transmission, the logical state is assumed to undergo ideal error correction, and the resulting logical error probability is evaluated analytically using Eq. \ref{t_qubit}. The simplified implementation of the non-local CNOT proceeds as follows:

\begin{enumerate}
    \item The travelling ancilla qubit \(B_1\) is transmitted through the optical fibre from the sender to the receiver. Amplitude damping is applied to this qubit. When cat-state encoding is employed, the physical damping channel is replaced by the equivalent logical phase-flip channel determined by equation (\ref{eq:alpha_pz}). The service provider is involved in the error correction process, as is each party.
    
    \item The sender (or controller) performs a local CNOT operation using its coin qubit as the control and the qubit of the shared Bell state.
    
    \item The Alice qubit of the Bell state is measured in the computational basis. To simplify the process, the measurement outcome is fixed to $\ket{0}$. In the present analysis, we consider the heralded successful branch in which the required measurement outcomes are zero. Trials yielding any other measurement outcome are discarded. The reported fidelity, therefore, corresponds to the conditional fidelity of a successfully heralded operation.
    
    \item The receiver performs a local CNOT operation using the received qubit of the Bell state as the control and the receiver's coin qubit as the target.
    
    \item Finally, the receiver applies a Hadamard gate to Bob's bell state qubit, which he later measures in the computational basis. Again, for computational simplification, the output is fixed to $\ket{0}$.
\end{enumerate}

This simplification in the simulation is the absence of feed-forward corrections, since the objective of the present work is to compare the influence of amplitude damping across different communication architectures, thereby fixing the measurement outcomes and significantly reducing the computational overhead of the simulations. For each protocol, the fibre length L was varied from 0 km to 50 km in 1 km increments along each branch of the star network. At every distance, the corresponding amplitude damping probability was calculated and applied to the travelling qubits according to the selected communication architecture. For n-qubit encoding, the corresponding t-qubit error correction was implemented at each node before transmitting it through the optical fibre. The resulting output density matrix was compared with the corresponding noiseless output using the Uhlmann fidelity. Repeating this procedure over the entire distance range produced the fidelity-distance curves presented in the following sections. The resulting fidelity values were then plotted as a function of transmission distance, enabling a direct comparison of the robustness of the four communication protocols against photon loss. In the entire simulation, no Gate error, memory decoherence error, or measurement error is considered.

\subsection{Fidelity vs Distance}
We investigate the effect of repetition-code-based error correction on the fidelity of the proposed controlled quantum communication protocol over a star quantum network, where each arm length is $L$ km. The transmission is simulated over an optical fibre with an attenuation coefficient of $\kappa=0.04605~\mathrm{km}^{-1}$, which gives the corresponding amplitude damping probability 
\begin{equation}
p=1-e^{-\kappa L}.
\end{equation}

\begin{figure}
\centering
\includegraphics[width=\columnwidth]{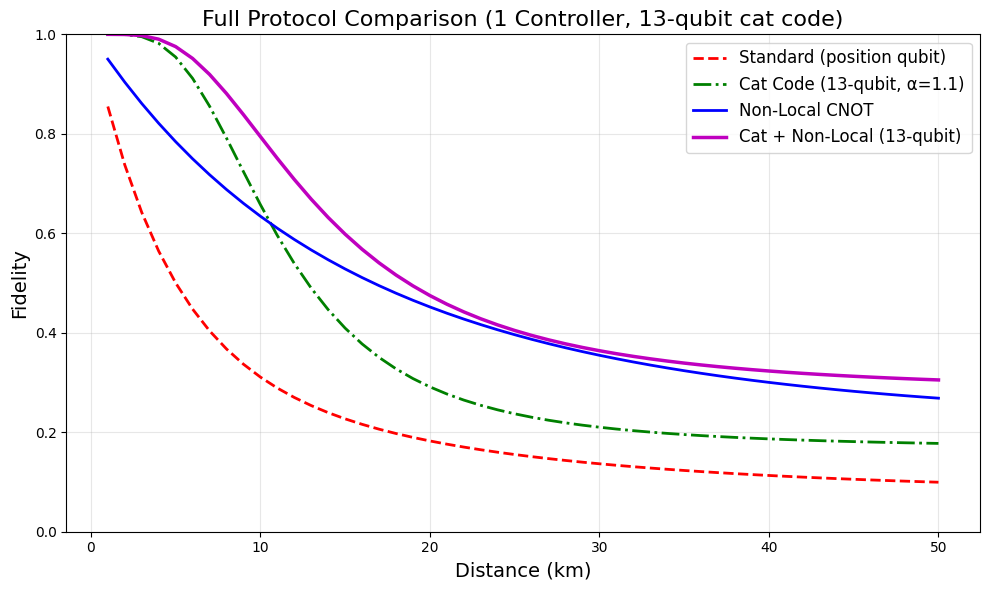}
\caption{Fidelity vs arm length distance in star network for four protocols: standard (red, dashed), cat code only (green, dash-dot), non-local CNOT (blue, solid), and cat code + non-local CNOT (magenta, solid). The cat-code-protected non-local CNOT protocol achieves the highest fidelity at all distances.}
\label{fig:fidelity_vs_distance}
\end{figure}

 For cat state $\alpha=1.1$, we plot the figure~\ref{fig:fidelity_vs_distance}, which shows the fidelity as a function of distance for all four protocols. The fidelities reported for the non-local CNOT architecture are conditional fidelities corresponding to the successfully heralded (0,0) measurement branch. We observe that the standard protocol suffers from rapid fidelity decay due to multiple amplitude damping events. The thirteen-qubit cat code improves the fidelity over the standard protocol. However, the improvement is limited because the position qubit still travels multiple times. The non-local CNOT protocol significantly outperforms both when the arm length increases beyond 10 km, as it eliminates the physical position qubit and reduces damping events to just two per EPR pair. The combined cat-code-protected non-local CNOT protocol, which uses a thirteen-qubit cat code to protect an EPR pair to perform non-local CNOT, achieves the highest fidelity, demonstrating the benefit of combining both techniques.

\subsection{Comparison of the standard protocol and repetition codes with increasing transmission distance}

\begin{figure}
    \centering
    \includegraphics[width=0.9\linewidth]{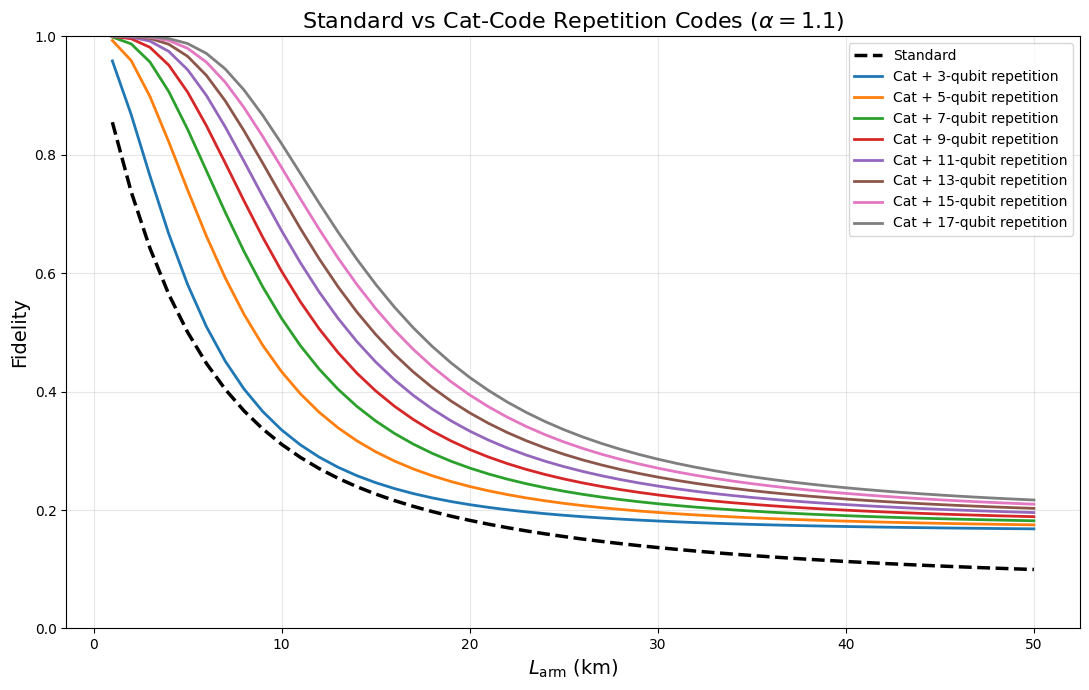}
    \caption{Comparison of the communication fidelity of the standard protocol and the proposed repetition-code-based schemes using 3-, 5-, 7-, 9-, 11-, 13-, 15- and 17-qubit repetition codes as a function of the optical fibre arm length of a star network for a single-controller star network. The cat-state amplitude is fixed at $\alpha=1.1$.}
    \label{fig:standard_vs_repetition}
\end{figure}

Figure~\ref{fig:standard_vs_repetition} compares the fidelity of the standard protocol (without repetition coding) with those employing  3-, 5-, 7-, 9-, 11-, 13-, 15- and 17-qubit repetition codes as a function of the transmission distance. As the transmission distance increases, photon loss accumulates, reducing fidelity. Increasing the repetition code length enables the correction of a larger number of logical phase-flip errors induced by amplitude damping, thereby improving the communication fidelity over longer distances. This improvement comes at the expense of additional physical qubits and increased error-correction overhead.

\subsection{Effect of controllers on the communication fidelity}

\begin{figure}
\centering

\begin{subfigure}[b]{0.45\textwidth}
    \centering
    \includegraphics[width=\linewidth]{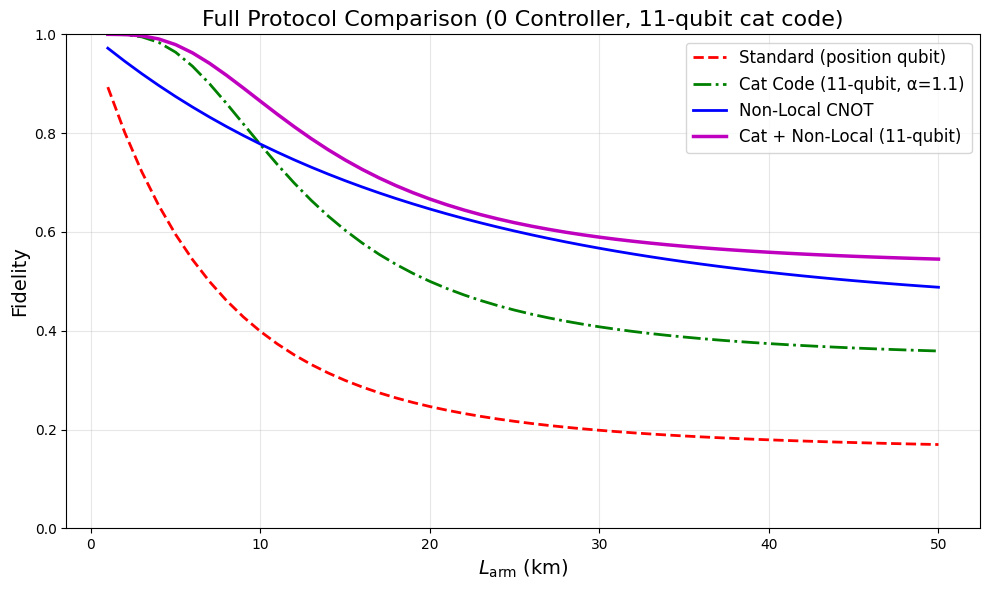}
    \caption{No controller.}
    \label{fig:no_controller}
\end{subfigure}
\hfill
\begin{subfigure}[b]{0.45\textwidth}
    \centering
    \includegraphics[width=\linewidth]{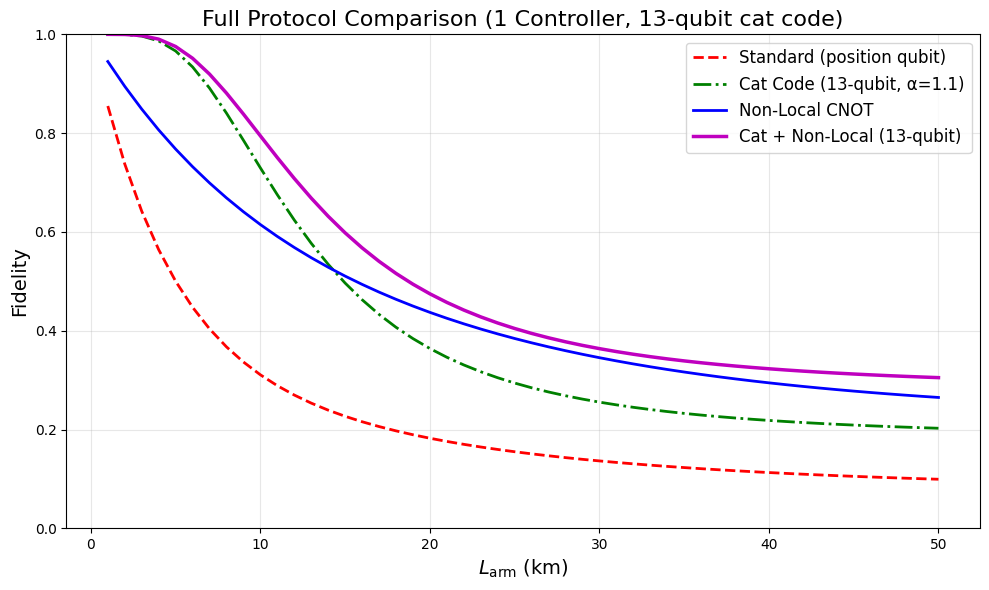}
    \caption{One controller.}
    \label{fig:one_controller}
\end{subfigure}
\hfill
\begin{subfigure}[b]{0.45\textwidth}
    \centering
    \includegraphics[width=\linewidth]{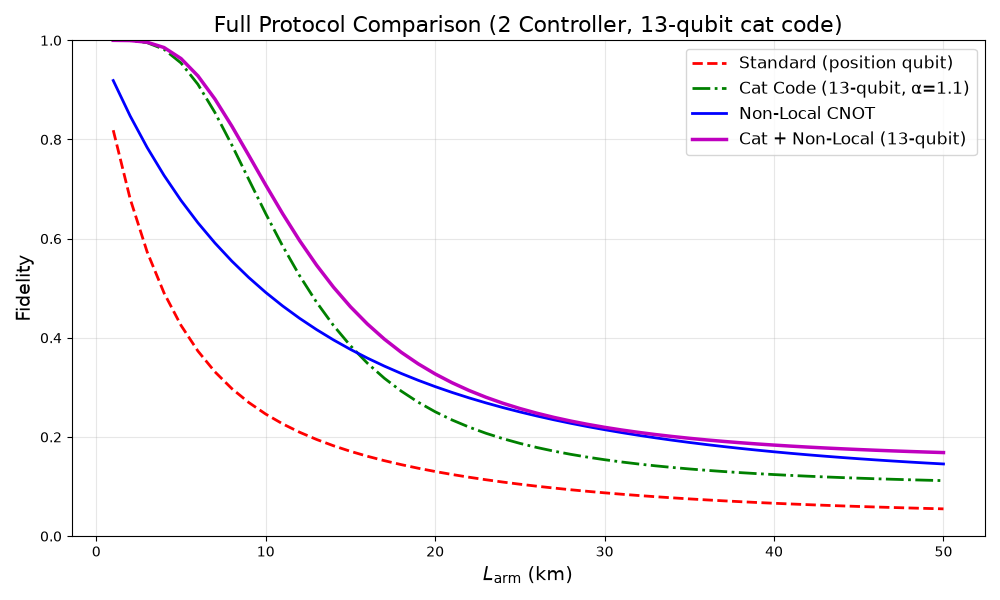}
    \caption{Two controllers.}
    \label{fig:two_controller}
\end{subfigure}

\caption{Comparison of the protocol fidelity on increasing the arm length of a star network for different numbers of controllers under amplitude damping. In each case, the cat-code-protected non-local CNOT protocol achieves the highest fidelity, whereas the standard protocol's performance deteriorates as the number of controllers increases.}
\label{fig:controller_comparison}
\end{figure}

The number of controllers directly affects the communication fidelity by increasing the number of optical-fibre transmissions required during the protocol, as shown in figure \ref{fig:controller_comparison}. In the standard implementation, each additional controller introduces two additional amplitude-damping events associated with the transmission of the position qubit. Consequently, the fidelity decreases rapidly with increasing controller count, dropping from 0.1698 at 50 km with no controllers to 0.0995 with one controller and 0.055 with two controllers.

Integrating a thirteen-qubit repetition code with a cat-qubit encoding of $\alpha=1.1$ to the position qubit improves the fidelity. This improvement is limited, as it still undergoes multiple optical-fibre transmissions and up to 6-phase flip error correction at each node, and the service provider also corrects only 6- phase flip errors. The higher-order error still adds up. At 50 km, the fidelity increases from 0.1698 to 0.367 for zero controllers, from 0.0995 to 0.2027 for one controller, and from 0.055 to  0.1120 for two controllers.

A significantly larger improvement is achieved by replacing the repeated transmission of the position qubit with optimal non-local CNOT gates without using cat qubits. Since only the distributed Bell pair is transmitted through the optical fibre, the number of amplitude-damping events is substantially reduced. As a result, the fidelity at 50 km increases to 0.4881, 0.2651, and 0.1456  for zero, one, and two controllers, respectively. For the non-local CNOT architectures, the reported values correspond to the conditional fidelity of a successfully heralded operation; therefore, the fidelity improvement should be interpreted together with the corresponding success probability.

The highest fidelities are obtained by combining the thirteen-qubit cat-code protection at each node and service centre with the non-local CNOT architecture. The corresponding fidelities at 50 km are 0.5524, 0.3052, and 0.1686 for zero, one, and two controllers, respectively. These results demonstrate that the non-local CNOT architecture provides the primary improvement by reducing photon-loss events. At the same time, the thirteen-qubit repetition code further suppresses the residual logical phase-flip errors. Together, they provide a scalable approach for maintaining higher fidelity in controlled quantum communication. 

\subsection{Effect of Cat State Amplitude on Fidelity}

\begin{figure}
    \centering
    \begin{subfigure}[b]{0.45\columnwidth}
        \centering
        \includegraphics[width=\linewidth]{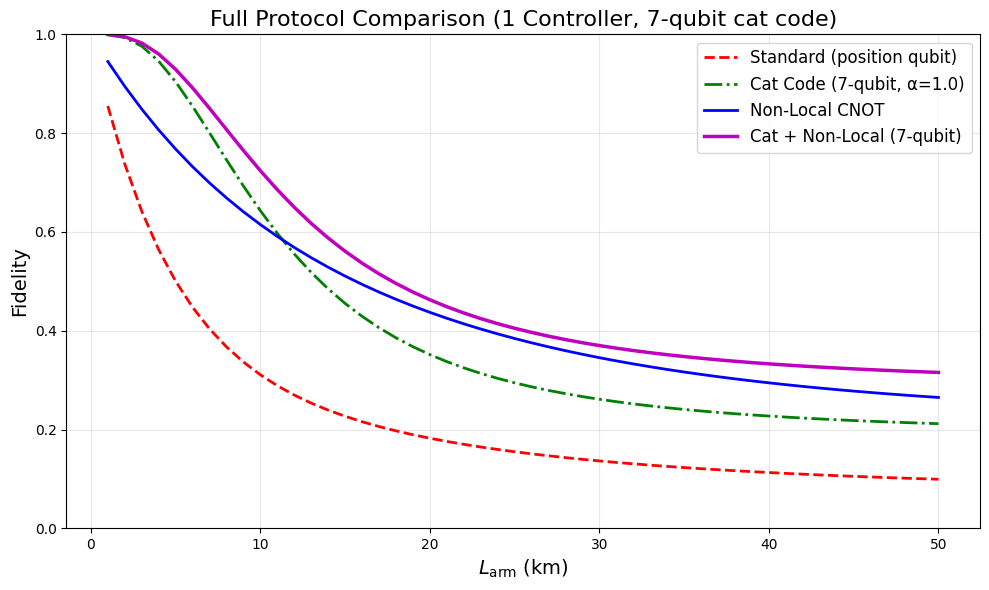}
        \caption{$\alpha = 1.0$}
        \label{fig:alpha_08}
    \end{subfigure}
    \begin{subfigure}[b]{0.45\columnwidth}
        \centering
        \includegraphics[width=\linewidth]{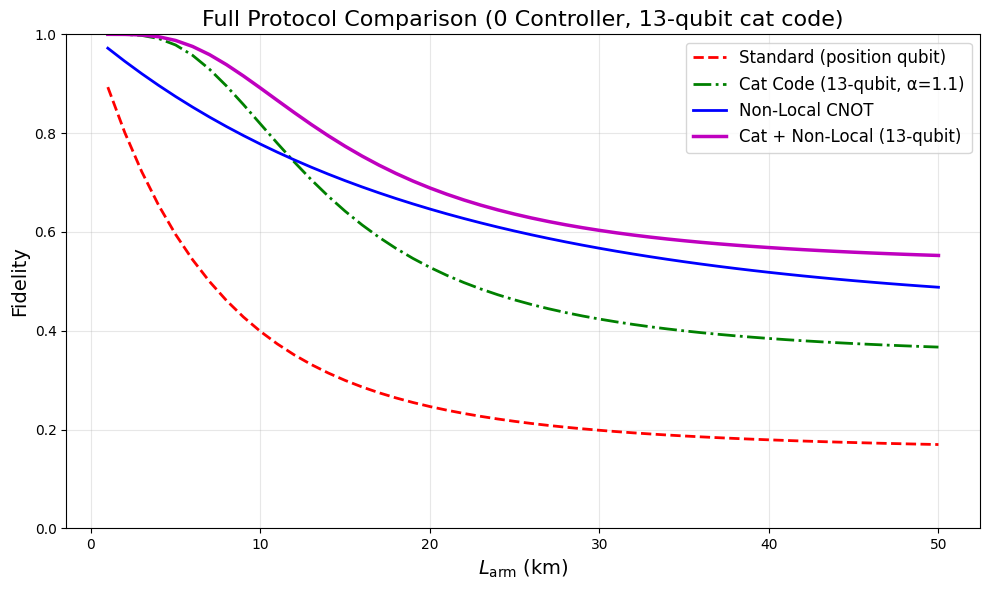}
        \caption{$\alpha = 1.1$}
        \label{fig:alpha_10}
    \end{subfigure}
    \begin{subfigure}[b]{0.45\columnwidth}
        \centering
        \includegraphics[width=\linewidth]{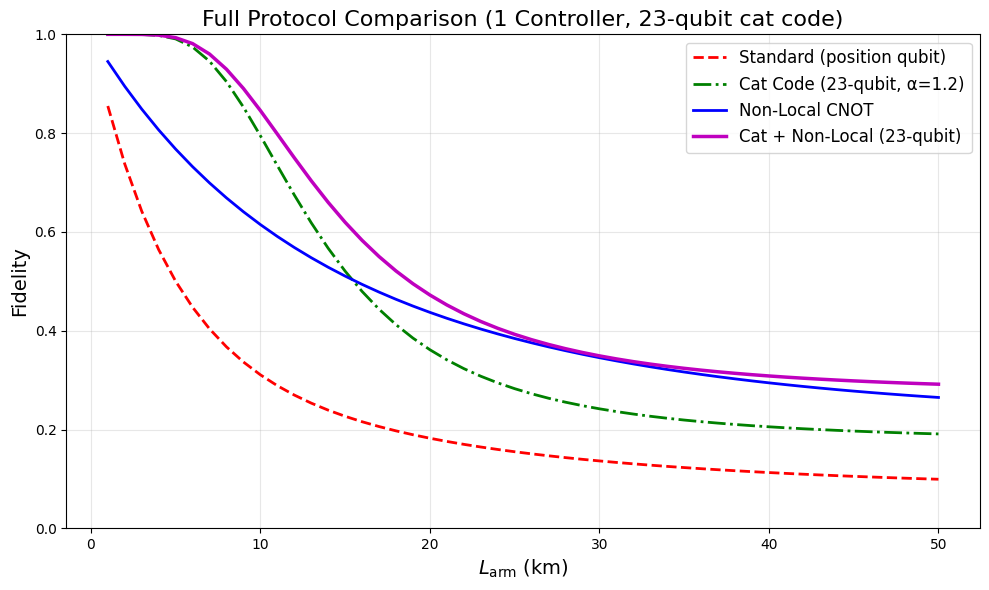}
        \caption{$\alpha = 1.2$}
        \label{fig:alpha_12}
    \end{subfigure}
    \caption{Fidelity vs arm length of a star network for four protocols with one controller for different cat state amplitudes: (a) $\alpha = 1.0 $, (b) $\alpha = 1.1$, and (c) $\alpha = 1.2$. }
    \label{fig:cat_alpha_comparison}
\end{figure}

The amplitude of the cat state plays a critical role in determining the protocol's fidelity. Increasing the cat state improves distinguishability. However, as the cat state's amplitude increases, the phase-flip probability increases. Thus, an optimal value that balances state distinguishability and phase-flip must be achieved. The cat state, $\alpha = 1.0$, is an improvement over the standard protocol with just a seven-qubit repetition code. However, the overlap between the two logical states is $0.135$. For the cat state, $\alpha = 1.1$ provides improvement over the standard protocol with an eleven-qubit repetition code. The overlap between the two logical states is $0.088$. For the cat state, $\alpha = 1.2$  improvement over the standard protocol with a twenty-three-qubit repetition code. The overlap between the two logical states is $0.056$. Fig.~\ref{fig:cat_alpha_comparison} shows the fidelity of their error-correcting code. 

\subsection{Effect of repetition code size}
\begin{figure}
    \centering
    \begin{subfigure}{0.45\columnwidth}
        \centering
        \includegraphics[width=\linewidth]{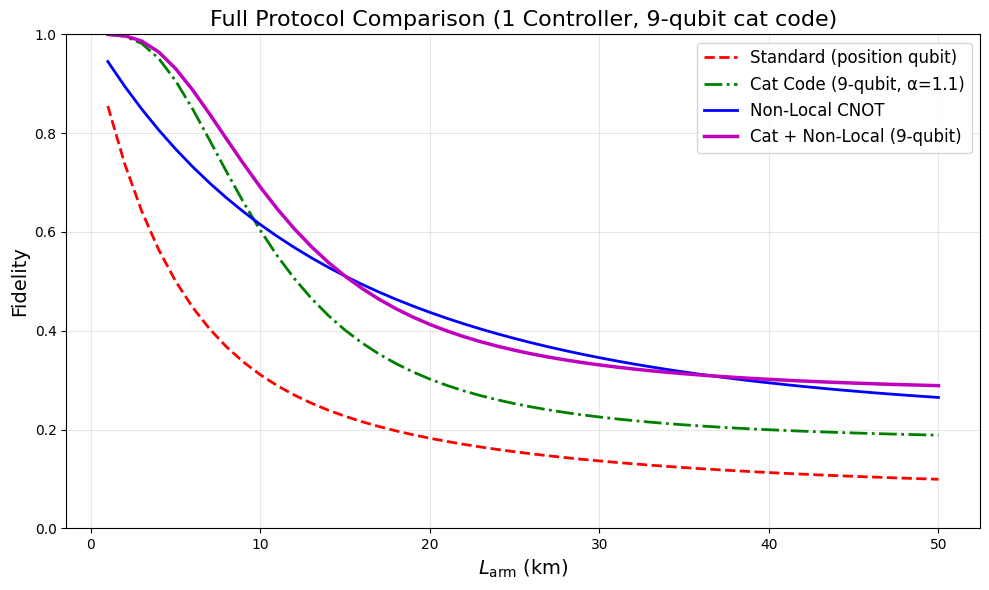}
        \caption{Nine-qubit repetition code}
        \label{fig:9qubit}
    \end{subfigure}
    \begin{subfigure}{0.45\columnwidth}
        \centering
        \includegraphics[width=\linewidth]{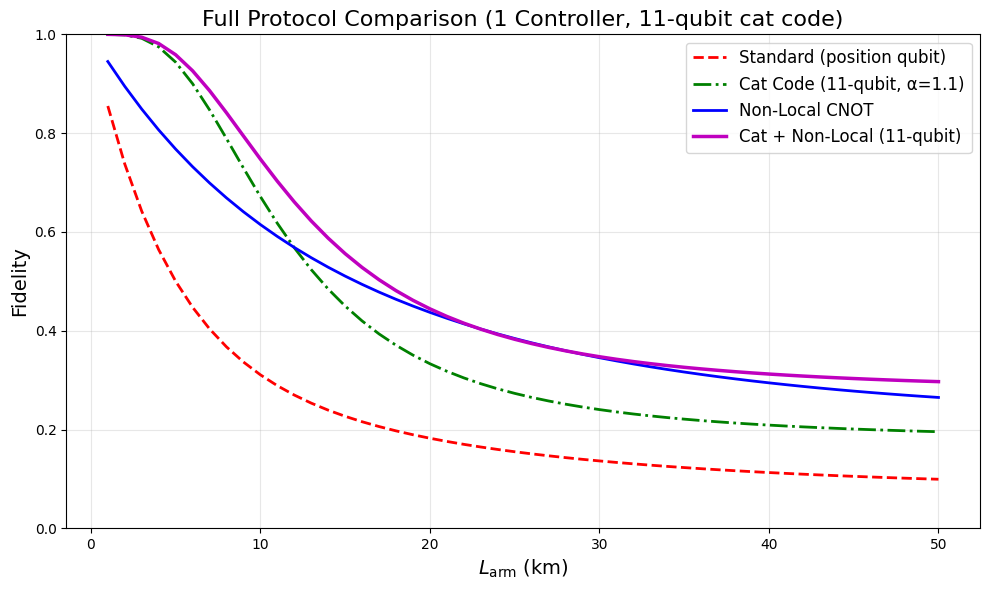}
        \caption{Eleven qubit repetition code}
        \label{fig:11qubit}
    \end{subfigure}
    \begin{subfigure}{0.45\columnwidth}
        \centering
        \includegraphics[width=\linewidth]{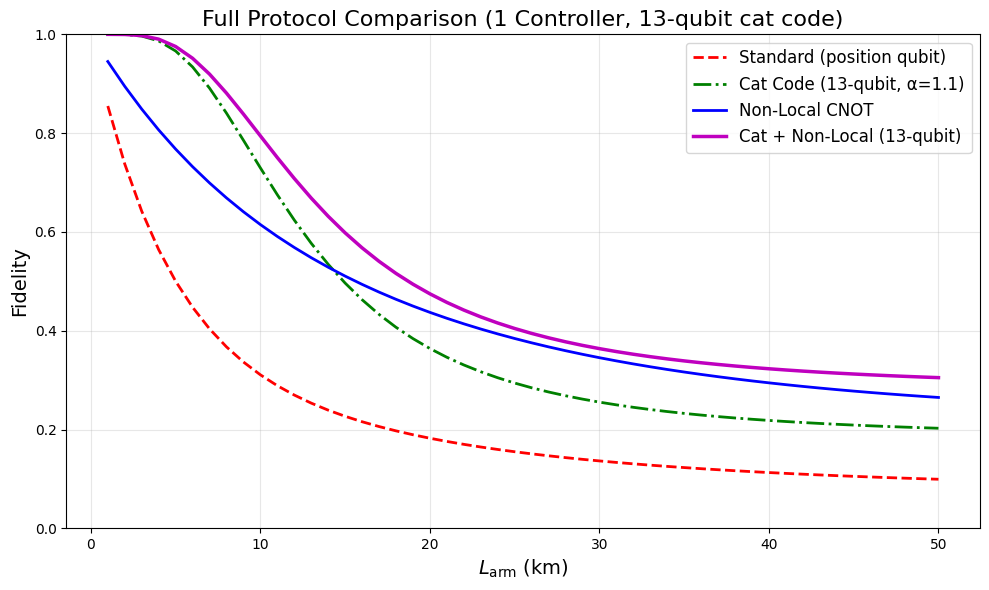}
        \caption{Thirteen qubit repetition code}
        \label{fig:13qubit}
    \end{subfigure}
    \begin{subfigure}{0.45\columnwidth}
        \centering
        \includegraphics[width=\linewidth]{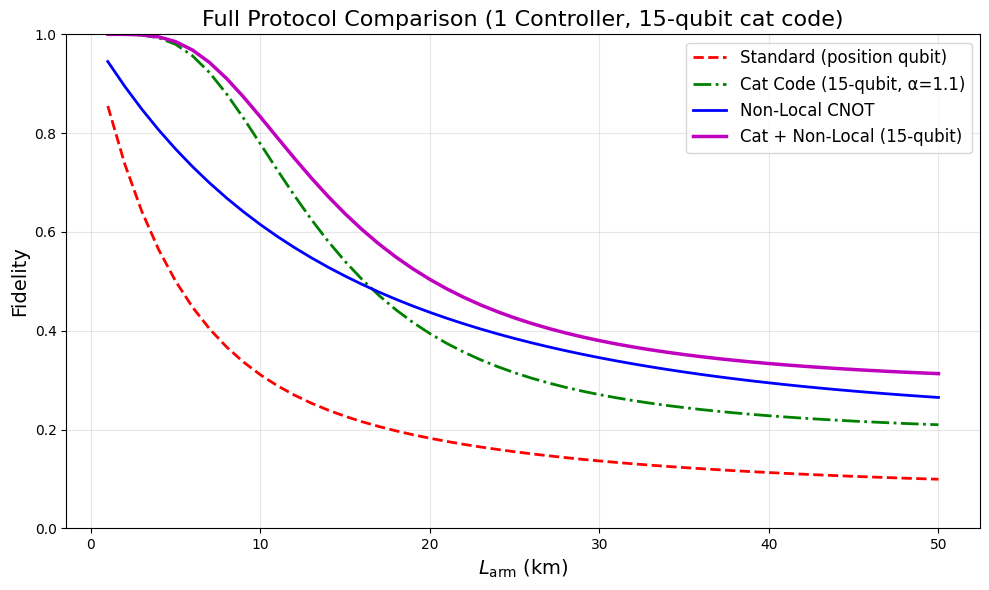}
        \caption{Fifteen qubit repetition code}
        \label{fig:15qubit}
    \end{subfigure}
    \caption{Fidelity vs arm length of the star network for the proposed protocol with one controller using 9-, 11-, 13-, and 15-qubit repetition codes. The cat-state amplitude is fixed at $\alpha=1.1$ to illustrate the effect of increasing the repetition code size.}
    \label{fig:increasingdepth}
\end{figure}

To understand the effect of increasing code size on fidelity. The simulation is performed for one controller with cat state $\alpha=1.1$. Increasing the size of the repetition code improves the correction of higher-photon-loss errors, as shown in equation \ref{t_qubit}. The figure \ref{fig:increasingdepth} shows that increasing code size generally improves state fidelity. For the cat state $\alpha=1.1$, each repetition codes outperforming the standard protocol. The pure cat-encoded state for each repetition at the initial small arm distance in the star network outperformed the non-local CNOT, but as the arm length increased, the non-local CNOT outperformed. Similarly, for the nine-qubit repetition code for the pure non-local CNOT protocol and the cat-encoded non-local CNOT protocol, there are two crossings of fidelity starting from around 15 km to 35 km, where the pure non-local CNOT outperforms the cat-encoded CNOT protocol, as seen in Figure \ref{fig:9qubit}. This occurs as phase flip degrades fidelity more rapidly than amplitude damping. However, this crossing over can be improved by increasing the repetition, as shown for the eleven-, thirteen-, and fifteen-qubit repetition codes in figures \ref{fig:11qubit}, \ref{fig:13qubit}, and \ref{fig:15qubit}. It can be seen that the pure non-local CNOT underperforms compared to the cat encoded non-local CNOT for the eleven-, thirteen-, and fifteen-qubit repetition codes. 

\section{Discussion and Conclusion\label{sec8}}

In this work, we presented an approach to implement a controlled quantum communication protocol in star quantum networks by combining optimal non-local CNOT gates with error-correcting cat-state encoding. It removes the need to repeatedly transport the position qubit through an optical fibre. Our results show that the non-local CNOT protocol increases the fidelity of the communication protocol. This improvement becomes increasingly significant as the number of controllers increases, because each additional controller contributes only two damping events on each bell state, rather than multiple damping events on the position qubit in the conventional protocol.

On the other hand, the repetition of the cat code alone provides comparatively less improvement when implemented in the standard protocol. Although it successfully converts amplitude damping into logical phase-flip errors, its error-correcting capability becomes insufficient once photon loss becomes significant, as correcting higher-order photon loss requires increasing no of qubits. So soon it becomes resource inefficient. In contrast, the thirteen-qubit cat code becomes highly effective when combined with the non-local CNOT protocol. Since the non-local implementation substantially suppresses amplitude damping, residual logical phase-flip errors persist and can be efficiently corrected by the cat code. This complementary behaviour produces the highest fidelity among all the protocols considered.

The security analysis also shows that the protocol using non-local CNOTs remains robust against beam-splitter-based eavesdropping attacks in noiseless settings. The CHSH inequality and the behaviour with the GHZ state change significantly when cat-encoded qubits are used. However, Eve's presence can be detected statistically. By deriving the corresponding detection threshold and applying repeated Bell tests, we show that an eavesdropper can be identified with arbitrarily high confidence.

Overall, this work provides a direction: integrating cat-state encoding with optimal non-local CNOT gates can significantly improve the performance of controlled quantum communication. These results represent an important step toward the realisation of secure, large-scale quantum communication networks.

\section{Limitations and Future works\label{sec9}}

Although the proposed protocol demonstrates a significant improvement in communication fidelity, several limitations remain. First, the present analysis assumes ideal quantum gates, perfect syndrome extraction, error-free correction, and ideal entanglement distribution while considering amplitude damping as the dominant source of noise. It also assumes quantum memories that perfectly preserve the encoded states throughout the communication process and remain coherent throughout the entire protocol. In practical implementations, gate imperfections, memory decoherence, dephasing, depolarisation, detector inefficiencies, and finite Bell-pair fidelities will further reduce the communication fidelity. Incorporating these realistic noise sources into the simulation framework would provide a more comprehensive evaluation of the protocol under experimentally relevant conditions. In addition, using these results as guidance, better-optimised, more qubit-efficient error-correcting codes should be developed.

The security analysis presented in this work focuses on beam-splitter attacks on the distributed entanglement resource. Although this attack captures an important vulnerability of cat-state-based optical communication, more sophisticated coherent, collective, and adaptive attacks remain to be investigated.

The proposed protocol can also be extended to more general network architectures to evaluate its scalability in distributed quantum communication. Furthermore, integrating entanglement purification and quantum repeater protocols with the proposed non-local CNOT architecture could improve the fidelity of the distributed Bell pairs before the communication protocol consumes them. Also, the cat qubit amplitude improvement. Finally, an experimental implementation using existing photonic quantum communication platforms would provide valuable validation of the theoretical predictions presented in this work and establish the practical feasibility of cat-code-protected controlled quantum communication over long-distance quantum networks.

\section{Author Contributions }

S.D. conceived the research problem, developed the theoretical framework,
performed the analytical calculations and numerical simulations, and wrote
the manuscript. A.S. and K.S. supervised the research, reviewed the
manuscript, provided overall guidance, and approved the final version.

\bibliographystyle{plainnat}
\bibliography{mdpi}

@article{PhysRevA.62.052317,
  title = {Optimal local implementation of nonlocal quantum gates},
  author = {Eisert, J. and Jacobs, K. and Papadopoulos, P. and Plenio, M. B.},
  journal = {Phys. Rev. A},
  volume = {62},
  issue = {5},
  pages = {052317},
  numpages = {7},
  year = {2000},
  month = {Oct},
  publisher = {American Physical Society},
  doi = {10.1103/PhysRevA.62.052317},
  url = {https://link.aps.org/doi/10.1103/PhysRevA.62.052317}
}

@article{Das2025,
  author    = {Subham Das and A. V. N. S. Meghnath and Rajiuddin Sk and Prasanta K. Panigrahi},
  title     = {Controlled quantum communication using quantum walk},
  journal   = {Quantum Information Processing},
  year      = {2025},
  volume    = {24},
  number    = {9},
  pages     = {271},
  doi       = {10.1007/s11128-025-04897-1},
  url       = {https://doi.org/10.1007/s11128-025-04897-1},
  issn      = {1573-1332}
}

@article{PhysRevA.81.062344,
  title = {Entanglement properties of optical coherent states under amplitude damping},
  author = {Wickert, Ricardo and Bernardes, Nadja Kolb and van Loock, Peter},
  journal = {Phys. Rev. A},
  volume = {81},
  issue = {6},
  pages = {062344},
  numpages = {6},
  year = {2010},
  month = {Jun},
  publisher = {American Physical Society},
  doi = {10.1103/PhysRevA.81.062344},
  url = {https://link.aps.org/doi/10.1103/PhysRevA.81.062344}
}

@article{Kimble2008,
  author    = {H. J. Kimble},
  title     = {The quantum internet},
  journal   = {Nature},
  year      = {2008},
  volume    = {453},
  number    = {7198},
  pages     = {1023--1030},
  doi       = {10.1038/nature07127},
  url       = {https://doi.org/10.1038/nature07127},
  issn      = {1476-4687}
}

@misc{zhang2023federatedlearningquantumsecure,
      title={Federated Learning with Quantum Secure Aggregation}, 
      author={Yichi Zhang and Chao Zhang and Cai Zhang and Lixin Fan and Bei Zeng and Qiang Yang},
      year={2023},
      eprint={2207.07444},
      archivePrefix={arXiv},
      primaryClass={quant-ph},
      url={https://arxiv.org/abs/2207.07444}, 
}

@article{Pan_2024,
   title={The Evolution of Quantum Secure Direct Communication: On the Road to the Qinternet},
   volume={26},
   ISSN={2373-745X},
   url={http://dx.doi.org/10.1109/COMST.2024.3367535},
   DOI={10.1109/comst.2024.3367535},
   number={3},
   journal={IEEE Communications Surveys \& Tutorials},
   publisher={Institute of Electrical and Electronics Engineers (IEEE)},
   author={Pan, Dong and Long, Gui-Lu and Yin, Liuguo and Sheng, Yu-Bo and Ruan, Dong and Ng, Soon Xin and Lu, Jianhua and Hanzo, Lajos},
   year={2024},
   pages={1898–1949} }

@article{SINGH2025100089,
title = {Advancements in secure quantum communication and robust key distribution techniques for cybersecurity applications},
journal = {Cyber Security and Applications},
volume = {3},
pages = {100089},
year = {2025},
issn = {2772-9184},
doi = {https://doi.org/10.1016/j.csa.2025.100089},
url = {https://www.sciencedirect.com/science/article/pii/S2772918425000062},
author = {Sunil K. Singh and Sudhakar Kumar and Anureet Chhabra and Akash Sharma and Varsha Arya and M. Srinivasan and Brij B. Gupta}
}

@article{Qin2009,
  author    = {SuJuan Qin and QiaoYan Wen and LuoMing Meng and FuChen Zhu},
  title     = {Quantum secure direct communication over the collective amplitude damping channel},
  journal   = {Science in China Series G: Physics, Mechanics and Astronomy},
  volume    = {52},
  number    = {8},
  pages      = {1208--1212},
  year      = {2009},
  month     = aug,
  doi       = {10.1007/s11433-009-0140-z},
  url       = {https://doi.org/10.1007/s11433-009-0140-z},
  issn      = {1862-2844}
}

@article{BENNETT20147,
title = {Quantum cryptography: Public key distribution and coin tossing},
journal = {Theoretical Computer Science},
volume = {560},
pages = {7-11},
year = {2014},
note = {Theoretical Aspects of Quantum Cryptography – celebrating 30 years of BB84},
issn = {0304-3975},
doi = {https://doi.org/10.1016/j.tcs.2014.05.025},
url = {https://www.sciencedirect.com/science/article/pii/S0304397514004241},
author = {Charles H. Bennett and Gilles Brassard}
}

@inproceedings{GuedesAssis2013,
  author    = {Elloá B. Guedes and Francisco M. de Assis},
  title     = {Quantum Key Distribution over Collective Amplitude Damping Quantum Channels},
  booktitle = {Proceedings of the Eighth International Multi-Conference on Computing in the Global Information Technology (ICCGI 2013)},
  year      = {2013},
  pages      = {271--276},
  publisher = {IARIA},
  isbn       = {978-1-61208-283-7},
  url        = {https://www.thinkmind.org/articles/iccgi_2013_13_30_10116.pdf},
}

@article{PhysRevLett.81.2594,
  title = {Decoherence-Free Subspaces for Quantum Computation},
  author = {Lidar, D. A. and Chuang, I. L. and Whaley, K. B.},
  journal = {Phys. Rev. Lett.},
  volume = {81},
  issue = {12},
  pages = {2594--2597},
  numpages = {0},
  year = {1998},
  month = {Sep},
  publisher = {American Physical Society},
  doi = {10.1103/PhysRevLett.81.2594},
  url = {https://link.aps.org/doi/10.1103/PhysRevLett.81.2594}
}

@article{PhysRevA.56.2567,
  title = {Approximate quantum error correction can lead to better codes},
  author = {Leung, Debbie W. and Nielsen, M. A. and Chuang, Isaac L. and Yamamoto, Yoshihisa},
  journal = {Phys. Rev. A},
  volume = {56},
  issue = {4},
  pages = {2567--2573},
  numpages = {0},
  year = {1997},
  month = {Oct},
  publisher = {American Physical Society},
  doi = {10.1103/PhysRevA.56.2567},
  url = {https://link.aps.org/doi/10.1103/PhysRevA.56.2567}
}

@article{smallestqcam,
  title = {Smallest quantum codes for amplitude-damping noise},
  author = {Dutta, Sourav and Jain, Aditya and Mandayam, Prabha},
  journal = {Phys. Rev. Res.},
  volume = {8},
  issue = {3},
  pages = {L032004},
  numpages = {6},
  year = {2026},
  month = {Jul},
  publisher = {American Physical Society},
  doi = {10.1103/6g5l-m4r1},
  url = {https://link.aps.org/doi/10.1103/6g5l-m4r1}
}

@article{PhysRevA.70.022317,
  title = {Transmission of optical coherent-state qubits},
  author = {Glancy, S. and Vasconcelos, H. M. and Ralph, T. C.},
  journal = {Phys. Rev. A},
  volume = {70},
  issue = {2},
  pages = {022317},
  numpages = {7},
  year = {2004},
  month = {Aug},
  publisher = {American Physical Society},
  doi = {10.1103/PhysRevA.70.022317},
  url = {https://link.aps.org/doi/10.1103/PhysRevA.70.022317}
}

@article{Ting_2005,
   title={Controlled quantum teleportation and secure direct communication},
   volume={14},
   ISSN={1741-4199},
   url={http://dx.doi.org/10.1088/1009-1963/14/5/006},
   DOI={10.1088/1009-1963/14/5/006},
   number={5},
   journal={Chinese Physics},
   publisher={IOP Publishing},
   author={Ting, Gao and Feng-Li, Yan and Zhi-Xi, Wang},
   year={2005},
   month=Apr, pages={893–897}
}

@article{Yan:24,
author = {Jieli Yan and Xiaoyu Zhou and Zhihui Yan and Xiaojun Jia},
journal = {Opt. Express},
number = {12},
pages = {21977--21987},
publisher = {Optica Publishing Group},
title = {Remote and controlled quantum teleportation network of the polarization squeezed state},
volume = {32},
month = {Jun},
year = {2024},
url = {https://opg.optica.org/oe/abstract.cfm?URI=oe-32-12-21977},
doi = {10.1364/OE.523111},
}

\onecolumn\newpage
\appendix

\appendix
\section{Appendix A}
\subsection{Optimal Non-Local CNOT Gates}

 A fundamental challenge is implementing a CNOT gate between qubits that belong to different parties. The standard approach, when implemented in a star network architecture for controlled quantum communication, uses a \emph{physical position qubit} that travels between nodes, accumulating amplitude damping at each transmission step. However, there exists an optimal way to implement a non-local CNOT using only local operations and classical communication (LOCC) \cite{PhysRevA.62.052317}, consuming minimal resources. We now describe this protocol, which forms the foundation of our distributed architecture.

Consider two parties, Alice (holding qubit $A$, the control) and Bob (holding qubit $B$, the target). They share an ancilla pair $A_1, B_1$ in the maximally entangled Bell state:

\begin{equation}
|\Phi^+\rangle_{A_1B_1} = \frac{1}{\sqrt{2}}(|00\rangle_{A_1B_1} + |11\rangle_{A_1B_1})
\end{equation}

Eisert et al.~\cite{PhysRevA.62.052317} showed that a CNOT gate between $A$ and $B$ can be implemented non-locally using a Bell pair and 2 classical bits. The protocol proceeds as follows: Alice applies a local CNOT with $A$ as control and $A_1$ as target. He measures $A_1$ in the computational basis and sends the result (1 bit) to Bob. If the result is $|1\rangle$, Bob applies an $X$ gate to $B_1$; otherwise, he does nothing.
 Then, Bob applies a local CNOT with $B_1$ as the control and $B$ as the target, followed by a Hadamard gate on $B_1$.
Then he measures $B_1$ in the computational basis and sends the result (1 bit) to Alice. If the result is $|1\rangle$, Alice applies a $Z$ gate to $A$; otherwise, she does nothing. At the end of this protocol, the gate $\text{CNOT}(A \rightarrow B)$ has been implemented, with no physical qubit travelling between the parties. The protocol consumes one shared Bell pair and requires two bits of classical communication, one transmitted in each direction.

\section{Appendix B}
\subsection{CHSH Test for the Cat-Encoded GHZ State\label{Sec5}}

We consider the three-partite entangled state
\begin{equation}
\ket{\Psi} = \frac{1}{\sqrt{2}} \left( \ket{0}_A \ket{\alpha}_B \ket{\beta}_C + \ket{1}_A \ket{-\alpha}_B \ket{-\beta}_C \right),
\end{equation}
where $\ket{\pm\alpha}_B$ and $\ket{\pm\beta}_C$ are coherent (or cat) states on qubits $B$ and $C$, respectively. The key feature of this encoding is that the logical basis states are not necessarily orthogonal. We define the overlaps
\begin{equation}
\mathcal{O}_B = \braket{-\alpha}{\alpha}, \qquad \mathcal{O}_C = \braket{-\beta}{\beta},
\end{equation}
which are generally nonzero.

\subsubsection{Correlation Function}

We compute the correlation function for measurements on qubits $A$ and $B$, while qubit $C$ is left unmeasured (identity operator):
\begin{equation}
E(a,b) = \bra{\Psi} (\vec{a}\cdot\vec{\sigma})_A \otimes (\vec{b}\cdot\vec{\sigma})_B \otimes I_C \ket{\Psi}.
\end{equation}

A direct calculation in the limit $\mathcal{O}_B \approx 0$ (where $\ket{\alpha}$ and $\ket{-\alpha}$ form an orthonormal basis) yields
\begin{equation}
\boxed{
E(a,b) = a_z b_z + \mathcal{O}_C (a_x b_x + a_y b_y)
}
\end{equation}
where $\vec{a} = (a_x, a_y, a_z)$ and $\vec{b} = (b_x, b_y, b_z)$ are unit vectors specifying the measurement directions, and $\mathcal{O}_C = \braket{-\beta}{\beta}$.

\subsubsection{Correlation Matrix}

The correlation function can be written in matrix form as $E(a,b) = \vec{a}^T T \vec{b}$, where the correlation matrix is
\begin{equation}
T = 
\begin{pmatrix}
\mathcal{O}_C & 0 & 0 \\
0 & -\mathcal{O}_C & 0 \\
0 & 0 & 1
\end{pmatrix}.
\end{equation}

\subsubsection{The CHSH Expression}

The CHSH expression is
\begin{equation}
S = E(a,b) + E(a,b') + E(a',b) - E(a',b').
\end{equation}

Substituting $E(a,b) = \vec{a}^T T \vec{b}$:
\begin{equation}
S = \vec{a}^T T \vec{b} + \vec{a}^T T \vec{b}' + \vec{a}'^T T \vec{b} - \vec{a}'^T T \vec{b}'.
\end{equation}

Factoring:
\begin{equation}
S = \vec{a}^T T (\vec{b} + \vec{b}') + \vec{a}'^T T (\vec{b} - \vec{b}').
\end{equation}

Define:
\begin{equation}
M = T = \text{diag}(\mathcal{O}_C, -\mathcal{O}_C, 1), \qquad
\vec{u} = \vec{b} + \vec{b}', \qquad
\vec{v} = \vec{b} - \vec{b}'.
\end{equation}

Then:
\begin{equation}
\boxed{
S = \vec{a}^T M \vec{u} + \vec{a}'^T M \vec{v}
}
\end{equation}

\subsubsection{Maximization Over $\vec{a}$ and $\vec{a}'$ Using Cauchy-Schwarz}

For fixed $\vec{b}$ and $\vec{b}'$, the vectors $\vec{u}$ and $\vec{v}$ are fixed. We maximize over the unit vectors $\vec{a}$ and $\vec{a}'$.

By the Cauchy-Schwarz inequality:
\begin{equation}
|\vec{a}^T M \vec{u}| \leq \|\vec{a}\| \|M \vec{u}\| = \|M \vec{u}\|,
\end{equation}
since $\|\vec{a}\| = 1$. Equality is achieved when $\vec{a}$ is parallel to $M \vec{u}$:
\begin{equation}
\vec{a} = \frac{M \vec{u}}{\|M \vec{u}\|}.
\end{equation}

Similarly:
\begin{equation}
|\vec{a}'^T M \vec{v}| \leq \|\vec{a}'\| \|M \vec{v}\| = \|M \vec{v}\|,
\end{equation}
with equality when:
\begin{equation}
\vec{a}' = \frac{M \vec{v}}{\|M \vec{v}\|}.
\end{equation}

Therefore, the maximum of $S$ over $\vec{a}$ and $\vec{a}'$ is:
\begin{equation}
\boxed{
S_{\text{max}} = \|M \vec{u}\| + \|M \vec{v}\|
}
\end{equation}

\subsubsection{Maximization Over $\vec{b}$ and $\vec{b}'$}

Since $\vec{b}$ and $\vec{b}'$ are unit vectors:
\begin{equation}
(\vec{b} + \vec{b}') \cdot (\vec{b} - \vec{b}') = \|\vec{b}\|^2 - \|\vec{b}'\|^2 = 0.
\end{equation}
Thus $\vec{u} = \vec{b} + \vec{b}'$ and $\vec{v} = \vec{b} - \vec{b}'$ are orthogonal. Also:
\begin{equation}
\|\vec{u}\|^2 + \|\vec{v}\|^2 = 2\|\vec{b}\|^2 + 2\|\vec{b}'\|^2 = 4.
\end{equation}

Since $\vec{u} \perp \vec{v}$, we can parameterize:
\begin{equation}
\vec{u} = 2\cos\theta \, \hat{n}_1, \qquad \vec{v} = 2\sin\theta \, \hat{n}_2,
\end{equation}
where $\hat{n}_1 \perp \hat{n}_2$ are unit vectors.

Then:
\begin{equation}
\|M\vec{u}\| = 2\cos\theta \|M\hat{n}_1\|, \qquad
\|M\vec{v}\| = 2\sin\theta \|M\hat{n}_2\|.
\end{equation}

Since $M = \text{diag}(\mathcal{O}_C, -\mathcal{O}_C, 1)$, the maximum of $\|M\hat{n}\|$ is $1$ (achieved when $\hat{n} = \hat{z}$), and the maximum under the constraint $\hat{n}_1 \perp \hat{n}_2$ is achieved by:
\begin{equation}
\hat{n}_1 = \hat{z}, \qquad \hat{n}_2 = \hat{x}.
\end{equation}

Then:
\begin{equation}
\|M\hat{n}_1\| = 1, \qquad \|M\hat{n}_2\| = \mathcal{O}_C.
\end{equation}

Therefore:
\begin{equation}
S_{\text{max}} = 2\cos\theta (1) + 2\sin\theta (\mathcal{O}_C) = 2(\cos\theta + \mathcal{O}_C \sin\theta).
\end{equation}

Maximizing over $\theta$:
\begin{equation}
\frac{d}{d\theta}(\cos\theta + \mathcal{O}_C \sin\theta) = -\sin\theta + \mathcal{O}_C \cos\theta = 0,
\end{equation}
which gives:
\begin{equation}
\tan\theta = \mathcal{O}_C.
\end{equation}

Thus:
\begin{equation}
\cos\theta = \frac{1}{\sqrt{1 + \mathcal{O}_C^2}}, \qquad
\sin\theta = \frac{\mathcal{O}_C}{\sqrt{1 + \mathcal{O}_C^2}}.
\end{equation}

Substituting:
\begin{equation}
S_{\text{max}} = 2\left( \frac{1}{\sqrt{1 + \mathcal{O}_C^2}} + \frac{\mathcal{O}_C^2}{\sqrt{1 + \mathcal{O}_C^2}} \right)
= 2\frac{1 + \mathcal{O}_C^2}{\sqrt{1 + \mathcal{O}_C^2}}
= 2\sqrt{1 + \mathcal{O}_C^2}.
\end{equation}

\subsubsection{Final Result}

The maximum CHSH value is therefore:
\begin{equation}
\boxed{
S_{\text{max}} = 2\sqrt{1 + \mathcal{O}_C^2}
}
\end{equation}

\subsubsection{Optimal Measurement Settings}

The optimal vectors that achieve this maximum are:
\begin{equation}
\begin{aligned}
\vec{a} &= (0, 0, 1), \\
\vec{a}' &= (1, 0, 0), \\
\vec{b} &= \frac{1}{\sqrt{1 + \mathcal{O}_C^2}} (\mathcal{O}_C, 0, 1), \\
\vec{b}' &= \frac{1}{\sqrt{1 + \mathcal{O}_C^2}} (-\mathcal{O}_C, 0, 1).
\end{aligned}
\end{equation}

\subsubsection{Special Cases}

\begin{itemize}
    \item If $\mathcal{O}_C = 0$ (orthogonal branches on qubit C): 
    \begin{equation}
    S_{\text{max}} = 2,
    \end{equation}
    which is the classical bound. No violation.
    
    \item If $\mathcal{O}_C = 1$ (identical branches on qubit C): 
    \begin{equation}
    S_{\text{max}} = 2\sqrt{2},
    \end{equation}
    which is the maximal Tsirelson bound.
    
    \item If $\mathcal{O}_C = e^{-2|\beta|^2}$ (coherent states):
    \begin{equation}
    S_{\text{max}} = 2\sqrt{1 + e^{-4|\beta|^2}}.
    \end{equation}
 
\end{itemize}

\subsubsection{CHSH Violation Condition}

The state violates the CHSH inequality whenever $S_{\text{max}} > 2$:
\begin{equation}
2\sqrt{1 + \mathcal{O}_C^2} > 2 \implies \mathcal{O}_C \neq 0.
\end{equation}

Thus, for coherent states where $\mathcal{O}_C = e^{-2|\beta|^2} > 0$ for any finite $\beta$, the state always violates the CHSH inequality, though the violation becomes exponentially small for large $\beta$.

\section{Appendix C}

\subsection{Statistical Detection Threshold}

For $N$ CHSH testing rounds, Alice and Bob can detect Eve only if the drop in the CHSH value is larger than the statistical uncertainty. Since each round is independent and has a statistical fluctuation $\epsilon_i$, we can denote each round as $S_1, S_2, S_3....S_{N_{exp}}$ whose values are given as
\begin{equation}
    S_i= 2\sqrt{2} + \epsilon_i
\end{equation}

Since $\epsilon_i$ is random for each process. Then, by the central limit theorem, the uncertainty in the estimated CHSH value scales as:

\begin{equation}
\delta = \frac{\sigma_0}{\sqrt{N_{exp}}} 
\end{equation}

Here $\sigma_0$ can be experimentally verified in the absence of Eve. We absorb the experimentally determined fluctuation scale into an effective number of testing rounds, defined by
\begin{equation}
    N_r \equiv \frac{N_{\mathrm{exp}}}{\sigma_0^2},
\end{equation}
so that
\begin{equation}
    \delta = \frac{1}{\sqrt{N_r}}.
\end{equation}

Thus, Eve is detected when:

\begin{equation}
S_{\text{max}} < 2\sqrt{2} - \frac{1}{\sqrt{N_r}}
\end{equation}

Substituting the CHSH expression from Eq.~\ref{eq:chsh_alpha}:

\begin{equation}
2\sqrt{1 + e^{-4x}} < 2\sqrt{2} - \frac{1}{\sqrt{N_r}}
\label{eq:detection_condition}
\end{equation}

We now solve this inequality for $x$. Divide both sides by 2:

\begin{equation}
\sqrt{1 + e^{-4x}} < \sqrt{2} - \frac{1}{2\sqrt{N_r}}
\end{equation}

For small deviations from the maximum violation, we have $x \ll 1$, so $e^{-4x} \approx 1 - 4x$. We expand the left-hand side around $e^{-4x} = 1$:

Let $y = e^{-4x}$. Then:

\begin{equation}
\sqrt{1 + y} = \sqrt{2} + \frac{1}{2\sqrt{2}}(y - 1) + \mathcal{O}((y-1)^2)
\end{equation}

Substituting $y = e^{-4x} \approx 1 - 4x$:

\begin{equation}
\sqrt{1 + e^{-4x}} \approx \sqrt{2} + \frac{1}{2\sqrt{2}}(-4x) = \sqrt{2} - \frac{2x}{\sqrt{2}} = \sqrt{2} - \sqrt{2}x
\end{equation}

Thus:

\begin{equation}
\boxed{\sqrt{1 + e^{-4x}} \approx \sqrt{2} - \sqrt{2}x}
\end{equation}

 Substitute into the inequality:

\begin{equation}
\sqrt{2} - \sqrt{2}x < \sqrt{2} - \frac{1}{2\sqrt{N_r}}
\end{equation}

Cancel $\sqrt{2}$ from both sides:

\begin{equation}
-\sqrt{2}x < -\frac{1}{2\sqrt{N_r}}
\end{equation}

Multiply by $-1$ (which reverses the inequality):

\begin{equation}
\sqrt{2}x > \frac{1}{2\sqrt{N_r}}
\end{equation}

 Solve for $x$:

\begin{equation}
x > \frac{1}{2\sqrt{2}\sqrt{N_r}}
\end{equation}

\begin{equation}
\boxed{x > \frac{1}{2\sqrt{2N_r}}}
\end{equation}

\medskip

Thus, the critical value of $x$ above which Eve is detected is:

\begin{equation}
\boxed{x_{\text{crit}} = \frac{1}{2\sqrt{2N_r}}}
\end{equation}
If Eve's tapping is greater than $x_{\text{crit}}$, an $N_{exp}$- round CHSH test will detect it.
\begin{table}[ht]
\centering
\caption{Detection regimes based on the value of $x$}
\label{tab:detection_regimes}
\begin{tabular}{c|c}
\hline
\textbf{Condition} & \textbf{Meaning} \\
\hline
$x < \dfrac{1}{2\sqrt{2N_r}}$ & Eve is \textbf{not detected}; she taps too little light to be statistically significant \\
\hline
$x > \dfrac{1}{2\sqrt{2N_r}}$ & Eve is \textbf{detected}; the drop in CHSH exceeds the statistical uncertainty \\
\hline
\end{tabular}
\end{table}

 Using Eq.~\ref{eq:x_from_epsilon}, Eve has distinguishability $\epsilon$ when:

\begin{equation}
x = -\frac{1}{4}\ln(1 - 4\epsilon^2)
\end{equation}

Eve is detected when:

\begin{equation}
-\frac{1}{4}\ln(1 - 4\epsilon^2) > \frac{1}{2\sqrt{2N_r}}
\end{equation}

Solving for $\epsilon$:

\begin{equation}
-\ln(1 - 4\epsilon^2) > \frac{2}{\sqrt{2N_r}} = \frac{\sqrt{2}}{\sqrt{N_r}}
\end{equation}

\begin{equation}
1 - 4\epsilon^2 < e^{-\sqrt{2/N_r}}
\end{equation}

\begin{equation}
\epsilon > \frac{1}{2}\sqrt{1 - e^{-\sqrt{2/N_r}}}
\end{equation}

For large $N_r$, $e^{-\sqrt{2/N_r}} \approx 1 - \sqrt{2/N_r}$, so:

\begin{equation}
\epsilon > \frac{1}{2}\sqrt{\frac{\sqrt{2}}{\sqrt{N_r}}} = \frac{\sqrt[4]{2}}{2\sqrt[4]{N_r}}
\label{threshold}
\end{equation}

Thus, when the distinguishability of the Eve state is beyond the threshold \ref{threshold}, the state will be 
\begin{equation}
\boxed{\epsilon > \frac{\sqrt[4]{2}}{2\sqrt[4]{N_r}}}
\end{equation}

\medskip

\noindent \textbf{Summary of Key Results:}

\begin{table}[ht]
\centering
\caption{Summary of detection thresholds}
\label{tab:detection_summary}
\begin{tabular}{c|c}
\hline
\textbf{Quantity} & \textbf{Expression} \\
\hline
Critical $x$ for detection & $x_{\text{crit}} = \dfrac{1}{2\sqrt{2N_r}}$ \\
Eve detected if & $\epsilon > \dfrac{\sqrt[4]{2}}{2\sqrt[4]{N_r}}$ \\
Required $N_r$ for given $\epsilon$ & $N_r > \dfrac{1}{2[-\ln(1 - 4\epsilon^2)]^2}$ \\
\hline
\end{tabular}
\end{table}

Eve's distinguishability not only quantifies her information gain but also provides an estimate of the coherent-state amplitude available in the tapped mode. The overlap between the coherent states $|\beta\rangle$ and $|-\beta\rangle$ is

\begin{equation}
\langle \beta | -\beta \rangle
=
e^{-2|\beta|^2},
\end{equation}

so that

\begin{equation}
\left|\langle \beta | -\beta \rangle\right|^2
=
e^{-4|\beta|^2}.
\label{eq:coherent_overlap}
\end{equation}

Using Eq.~\eqref{eq:epsilon_def}, we obtain

\begin{equation}
2\epsilon
=
\sqrt{1-e^{-4|\beta|^2}},
\end{equation}

which immediately gives

\begin{equation}
\boxed{
\left|\langle \beta | -\beta \rangle\right|
=
\sqrt{1-4\epsilon^2}.
}
\label{eq:overlap_from_epsilon}
\end{equation}

Consequently, the coherent-state amplitude can be estimated directly from Eve's distinguishability as

\begin{equation}
\boxed{
|\beta|^2
=
-\frac{1}{4}
\ln\!\left(1-4\epsilon^2\right).
}
\end{equation}

If Eve taps the communication channel using a beam splitter of transmissivity $T$, then the tapped coherent-state amplitude satisfies.

\begin{equation}
|\beta|^2=(1-T)|\alpha|^2,
\end{equation}

where $|\alpha\rangle$ is the coherent state transmitted by Alice. Therefore, measuring the distinguishability $\epsilon$ allows Eve to estimate both the overlap of the intercepted coherent states and the fraction of optical power extracted from the communication channel.
Using Eq ~\eqref{abc} with Eq.~\eqref{eq:beta_from_epsilon}, we obtain

\begin{equation}
(1-T)|\alpha|^2
=
-\frac14
\ln\!\left(1-4\epsilon^2\right).
\end{equation}

Hence, the channel transmissivity can be estimated as

\begin{equation}
\boxed{
T
=
1+
\frac{1}{4|\alpha|^2}
\ln\!\left(1-4\epsilon^2\right)
=
1-
\frac{-\ln\!\left(1-4\epsilon^2\right)}
{4|\alpha|^2}.
}
\end{equation}

\end{document}